\documentclass[apj]{emulateapj}
\usepackage{natbib}
\usepackage{multirow}

\begin{document}

\title{The dependence of stellar mass and angular momentum losses on
  latitude and on active region and dipolar magnetic fields}

\author{ Cecilia Garraffo\altaffilmark{1},  Jeremy
  J. Drake\altaffilmark{1},  Ofer Cohen\altaffilmark{1}}

\altaffiltext{1}{Harvard-Smithsonian Center for Astrophysics, 60 Garden St. Cambridge, MA 02138}

\begin{abstract}

Rotation evolution of late-type stars is dominated by magnetic
braking and the underlying factors that control this angular
momentum loss are important for the study of stellar spin-down.  
In this work, we study angular momentum loss as a function of two different aspects of magnetic
activity using a calibrated Alfv\'en wave-driven magnetohydrodynamic wind model: the strengths of magnetic spots and their distribution in latitude.  By driving the model using solar and modified solar surface magnetograms, 
we show that the topology of the field arising from the net interaction of both small-scale and large-scale field is 
important for spin-down rates and that angular momentum loss is not a simple
function of large scale magnetic field strength.  We find that changing the latitude of magnetic spots can modify mass and angular momentum loss rates by a factor of two.    The general effect that causes these differences is the closing down of large-scale open field at mid- and high-latitudes by the addition of the small-scale field.  These effects might give rise to modulation of mass and angular momentum loss through stellar cycles, and present a problem for {\it ab initio} attempts to predict stellar spin-down based on wind models.  For all the magnetogram cases considered here, from dipoles to various spotted distributions, we find that angular momentum loss is dominated by the mass loss at mid-latitudes.  The spin-down torque applied by magnetized winds therefore acts at specific latitudes and is not evenly distributed over the stellar surface, though this aspect is unlikely to be important for understanding spin-down and surface flows on stars.

\end{abstract}

\keywords{stars: rotation - stars: magnetic field - stars: coronae -}

\section{INTRODUCTION}
\label{sec:Intro}

Stellar rotation and its changes over time are important for
understanding stellar evolution and many astrophysical phenomena of
interest, such as magnetic activity, close binaries, cataclysmic variables and mass transfer systems.

Stars appear to be born with a wide range of angular momenta but eventually converge to a rotation period
determined fairly uniquely by their age and spectral type, following the empirical  law $ \Omega
\sim t^{-1/2}$ \citep{Skumanich72}.  This singular evolutionary path provides the basis for ``gyrochronology'', using stellar rotation rate as a diagnostic of age \citep{Kawaler:89,Barnes03, Bouvier13, Allain96, Gallet13}.  At very early stages of evolution, disks are thought to help prevent contracting pre-main-sequence  stars from spinning
up \citep[e.g.][]{Camenzind90, Koenigl91,Rebull.etal:04}, although the theory is not without some difficulties \citep[e.g.][]{Matt.etal:04,Matt.etal:10}.  Once the
disk is dispersed, stars can spin up as they continue to contract until they reach the Zero Age
Main Sequence.
From then on
stars begin their spin-down process through magnetized winds that carry away plasma
and angular momentum, so called magnetic braking \citep{Weber67,
  Mestel68}.  The plasma in the winds escape their host star at a
certain distance at which the winds reach the Alfv\'enic speed
$u_A=B/\sqrt{4\pi \rho}$, with $B$ being the local magnetic field strength, and $\rho$ the local mass density.  The collection of points where this happens
forms the Alfv\'en surface and can be thought of as a boundary
outside of which events are magnetically disconnected from the star.

The Skumanich law suggests that angular momentum loss in young stars is determined by their
rotation period, being more efficient for faster rotating stars.  Dynamo
processes are believed to scale with rotation, providing stronger magnetic fields and, it is assumed,
stronger winds for faster rotation
($B_0 \sim \Omega^2 $)  \citep[see][and references therein]{Jeffries14}. 
Quantitative details of this mechanism are uncertain due to the lack of
direct wind measurements in main-sequence stars
with outer convective envelopes (except for our own Sun).  Several different techniques have been developed so far to indirectly constraint stellar mass
loss rates.  Some of these place upper limits using infrared and radio observations
\citep{Cohen82, Hollis85, Lim96,
Gaidos00}.  Others measure chemical separation and abundances of
metals \citep{Michaud86}, $H\alpha$ line profiles \citep{Lanz92, Bertin95}, and $Ly\alpha$ absorption at
the limit of the termination shock into the ISM \citep[the ``astrosphere''][]{Wood02, Wood05,
  Wood06, Wood14}. 
However, these techniques are either fairly uncertain or else rely on assumptions about unknown quantities,
such as the interstellar-medium conditions at the boundary of the astrosphere.

The first analytical
expression relating mass and angular momentum loss was given by
\cite{Weber67}, 
\begin{equation}
\dot{J}=\frac{2}{3} \Omega \dot{M}R_A^2 ,
\label{WD}
\end{equation}
where $\dot{J}$ is the angular momentum loss rate, $\Omega$ the
angular velocity, $\dot{M}$ the mass loss rate, and
where a constant radial field was assumed at the surface of the star.  While a good 
starting point for calculating stellar angular momentum loss, this prescription breaks down for
more complex magnetic topologies.  Stellar magnetic field geometries are difficult to measure
reliably but are known to be far from that
of a split monopole.  Zeeman Doppler Imaging of stars \citep{Semel80,Donati97} has provided magnetic maps for
nearly a hundred stars so far \citep[see, for example,][and references
therein]{Donati09, Morin10}, albeit with very limited spatial resolution.  The simplest of these maps are close to dipolar fields,
for which the Weber and Davis equation already gives an overestimation of
angular momentum loss \citep{Kawaler88}. Young
and fast rotating stars, in particular, often show more complex field 
configurations \citep[e.g.][]{Donati.Landstreet:09}.

Despite the evidence for significant higher order magnetic field configurations on active stars, progress in understanding the effects of complex magnetic geometry on angular momentum loss has been limited.  Analytical models have been built aiming to generalize the Weber and Davis angular momentum loss equation to more
realistic scenarios \citep{Mestel87, Kawaler88}, and including saturation effects \citep[e.g.][]{Chaboyer95a}.  However, 
these studies still probed only a very narrow range of low-order field geometries between radial and dipole configurations, and generally assumed a spherically-symmetric wind. 

Some authors have suggested that more complex magnetic
topology might play an important role on stellar spin-down. 
\cite{Holzwarth05a} and \cite{Holzwarth05b}  discussed the effect of 
non-uniform flux distributions on stellar winds and spin-down rates based on a polytropic wind model and concluded that that latitude of magnetic flux represents an additional variable in determining stellar winds and rotational evolution.
\cite{Garraffo13} has shown that small scale magnetic topology has an
impact on the relative number of open and closed field lines which
affects the wind-driven mass loss.  \cite{Lang14} have also concluded that small scale magnetic activity
significantly affects the closed loop structure.  This will, in turn,
affect remaining open field lines (for a constant magnetic energy) that carry
the mass being lost to the wind.  \cite{Vidotto10} explored the effect on wind structure
of a misalignment between the axis of rotation and the magnetic axis
of a star.  \cite{Vidotto11, Vidotto12} simulated the winds of
observed ZDI maps and found a factor of two variation in angular
momentum loss rates over the magnetic cycle of $\tau$ Boo.
More recently, \cite{Vidotto14b} suggested
that the large scatter in the correlation between rotation period and
magnetic field strength might be due to different field geometries.  It
is worth mentioning that after some X-ray monitoring
\cite{Poppenhaeger12} found no long-term X-ray modulation of the
magnetic cycle of $\tau$ Boo.  

Recently, \citet{Reville.etal:15} have
modeled the effects of different magnetic topologies on the angular
momentum loss rates in a 2.5D cylindrical axisymmetric
set-up, using a Parker wind that is imposed on the
magnetic field.  They found that more complex fields
lead to less torque. 
\citet{Garraffo.etal:15} used a 3D MHD model, that includes a physics
based, self-consistent acceleration of the wind from the
chromosphere (see Section~\ref{sec:Methods} for a more detailed description
of this method), to model the steady-state wind structures for
the first ten terms in the multipolar expansion of the stellar surface
magnetic fields.  They conclude that magnetic complexity can lead to
a rapid suppression of mass and angular momentum loss rates, and
can provide a simple explanation for the bimodal rotation of young, rapid
rotators that is compatible with observations of greater magnetic
complexity observed for young stars  \citep[e.g.][]{Donati03,
Donati09, Marsden11a, Waite11, Waite15}.

Numerical simulations have recently
made considerable progress as a tool to understand the interplay
between angular momentum loss and stellar surface magnetic field distribution. 
Using a thermally-driven, spherically symmetric wind model motivated
by the solar wind solution of \cite{Parker58}, \citet{Matt12}
performed a number of two-dimensional, axisymmetric
magnetohydrodynamic (MHD) simulations for different topologies and derived a set of 
semi-analytical scaling laws for the angular momentum loss as a function of rotation
period and a single-valued surface magnetic field
strength.  In this context, \cite{Matt.etal:15} and \citet{Johnstone15} used observed
rotational evolution as constraints on rotational models to derive
equations for wind mass loss as a function of stellar mass and
rotation rate.
 \cite{Cranmer11} developed
a model of the steady mass loss rates of  both main-sequence and evolved cool stars that is in
good agreement with observations (although for main-sequence stars such data are relatively scant). In their
model, the mass loss rates depend on the filling factor that partially encodes the
magnetic topology of the surface field.
Alternatively, \cite{Cohen09} used a three-dimensional MHD model to study the effect of magnetic active
regions on mass and angular momentum loss rates. They concluded that
that stellar spin down can be significantly dependent on the latitude of magnetic 
spots.  This effect arises largely because magnetic spots give rise to small-scale closed loops that are effective in eliminating open field that carries the stellar wind.

\citet{Garraffo.etal:15} concluded that
the magnetic multipole order of the surface magnetic field of stars
can be a crucial factor in mass and angular-momentum loss
rates.  In this paper, we study if the belts of active regions could also
influence the spin down of stars in a way that is not previously taken into account.

The potential for spots to close down open field lines raises the question of from what latitudes the dominant mass loss occurs. Mass loss predominantly at polar regions, for example, would be expected to result in lower angular momentum loss than equatorial mass loss, affecting spin down in ways that could not be accounted for in earlier approaches.  
In this work, we study the role of different magnetic topologies on 
angular momentum loss rates and investigate the latitudes at which the mass and angular momentum losses occur. 
In particular, we are interested in active latitudes and the rings of magnetic spots, combined with an underlying dipole, like we see in our Sun.  We explore the implications on the latitudinal dependence of mass and angular momentum losses of
changing the latitude of the spots, as well as the strengths of both
the underlying dipole and of the active regions.  We adopt a numerical MHD approach using a sophisticated and detailed code routinely applied to studies of the solar wind and coronal mass ejections.

This paper is organized as follows.  We describe the numerical methods in Section~\ref{sec:Methods}.  In
Section~\ref{sec:Results} we present our results.  Our main findings
and their implications are discussed in
Section~\ref{sec:Discussion}, and the results
are summarized in Section~\ref{sec:Conclusions}.


\section{NUMERICAL SIMULATION}
\label{sec:Methods}

\subsection{MHD model} 
\label{sec:Model}

The numerical approach we adopt employs realistic magnetograms as drivers of a model to predict the magnetospheric structure and wind characteristics.
In order to obtain this solution for the stellar corona and stellar wind,
we use the generic {\it BATS-R-US} code \citep{Powell99,Toth12}. The
code solves the set of MHD equations for the
conservation of mass, momentum, magnetic induction, and energy on a
spherical grid, logarithmic in $\hat{r}$. To better resolve
current sheets that form during the simulation the 
Adaptive Mesh Refinement (AMR) capabilities of the code were used.  This feature automatically refines the grid around regions where the magnetic field changes its sign. 

The {\it BATS-R-US} module that models the solar (or stellar) corona is
driven by synoptic maps of the radial stellar magnetic field which
are used to specify the magnetic
field boundary conditions. Only the
potential component of the field influences the wind structure
\citep[see][]{Jardine13}.  Since surface meridional and azimuthal components of the field can quite accurately be predicted from the radial component in the potential field model, it is possible to drive a wind using just the radial component at the surface as input. The initial condition for the three-dimensional field is
obtained by calculating the potential field of these boundary
conditions assuming that the field is purely radial at a distance of
$r=4.5R_\star$ (the ``source surface"; \citealt{Altschuler67}).  In the case of the Sun,
the location of the source surface can be closer in than $4.5R_\odot$. We use
this larger value to prevent this parameter having any influence on the solution.  For strong
stellar magnetic fields, choosing a small value for the source surface
may truncate the magnetic loops resulting in an overpowering of the
stellar wind to unrealistically high values. This effect is eliminated as the
source surface is moved outwards to larger distances. 

Much existing MHD modeling of stellar winds, including some applications of  the {\it BATS-R-US} code, assumes a preexisting hot corona and implicitly accelerated wind by imposing an isotropic thermally-driven (Parker) wind \citep[e.g.][]{Matt12,Vidotto14b,Reville.etal:15}.  In contrast to those methods, we use a much more comprehensive and rigorous approach that accounts for the coronal heating and stellar wind acceleration self-consistently by employing detailed and tested physics as operating in the solar corona. This includes, on top of natural thermal acceleration due to the thermal pressure gradient, the dissipation of energy originating from Alfv\'en waves that are propagating along the magnetic loops. The model solves two additional energy equations for two counter-propagating Alfv\'en waves, which interact along the loop and drive a turbulent cascade. This cascade results in energy dissipation which is included in the MHD energy equation, and an additional momentum due to the Alfv\'en wave pressure gradient, which is included in the MHD momentum equation. In the open field lines, where only one wave originates from the field-line footpoint, the model employs a reflection coefficient that generates a counter-propagating wave so the turbulent cascade is generated along the open field lines as well. Finally, the model also accounts for electron heat conduction and radiative cooling in the MHD energy equation. A detailed description of all these physics-based terms and how they are obtained can be found in \cite{Oran13}, \cite{Sokolov13}, and in particular, in \cite{Vanderholst14}.

The wind in our model evolves with the
solution and the impact of the magnetic field topology on the wind is
much more significant, as is observed in the heliosphere
\citep{Phillips95,McComas07}. Not only our model accounts for the physical acceleration of
  the wind, but it also accounts for the different physical mechanisms
  that accelerate the fast and the slow solar (and stellar) wind. The mass loss rate
  is not affected directly by the redistribution of the magnetic field,
  but by the different acceleration applied on the wind plasma when
  the field topology is changed.  The response of the
  topology itself to this driving is also accounted for
  self-consistently.  These are fundamental
  differences between the approach adopted here and the approaches
  adopted in more isotropic models.

One of the advantages of the new model is that the base density is uniform and the density modulations are generated
self-consistently as a result of the physical coronal heating which is
implemented in the model. In contrast, the old version of the code \citep{Cohen07} used
an artificial scaling of the base density with the field line expansion
factor, which was designed to provide a solution that
compares better with observations of the solar wind. 
Furthermore, while both the old and the
new models compare well with solar wind data,
the new model compares much better with coronal global observations of
temperature and densities.

\subsection{Simulations}

\begin{figure*}
\includegraphics[width =\textwidth]{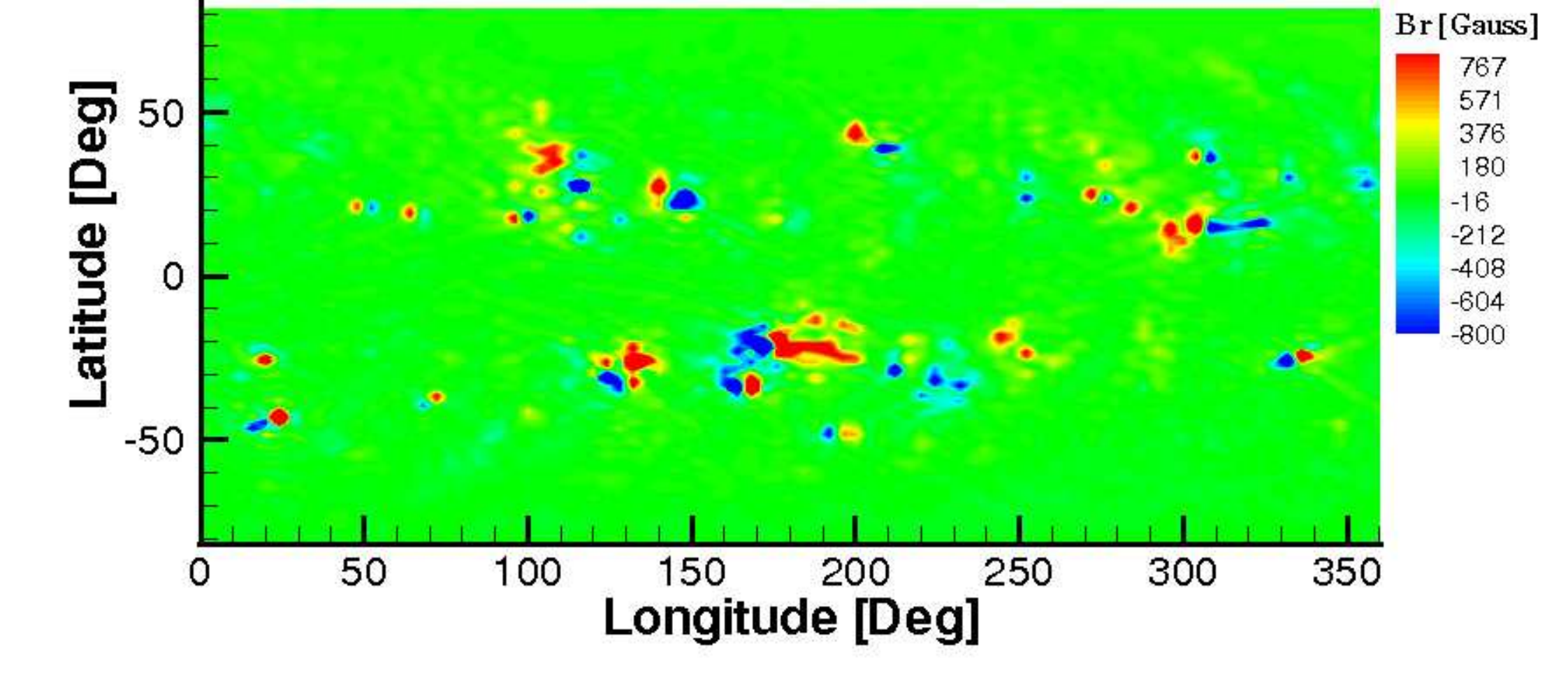}\\
\includegraphics[width = \textwidth]{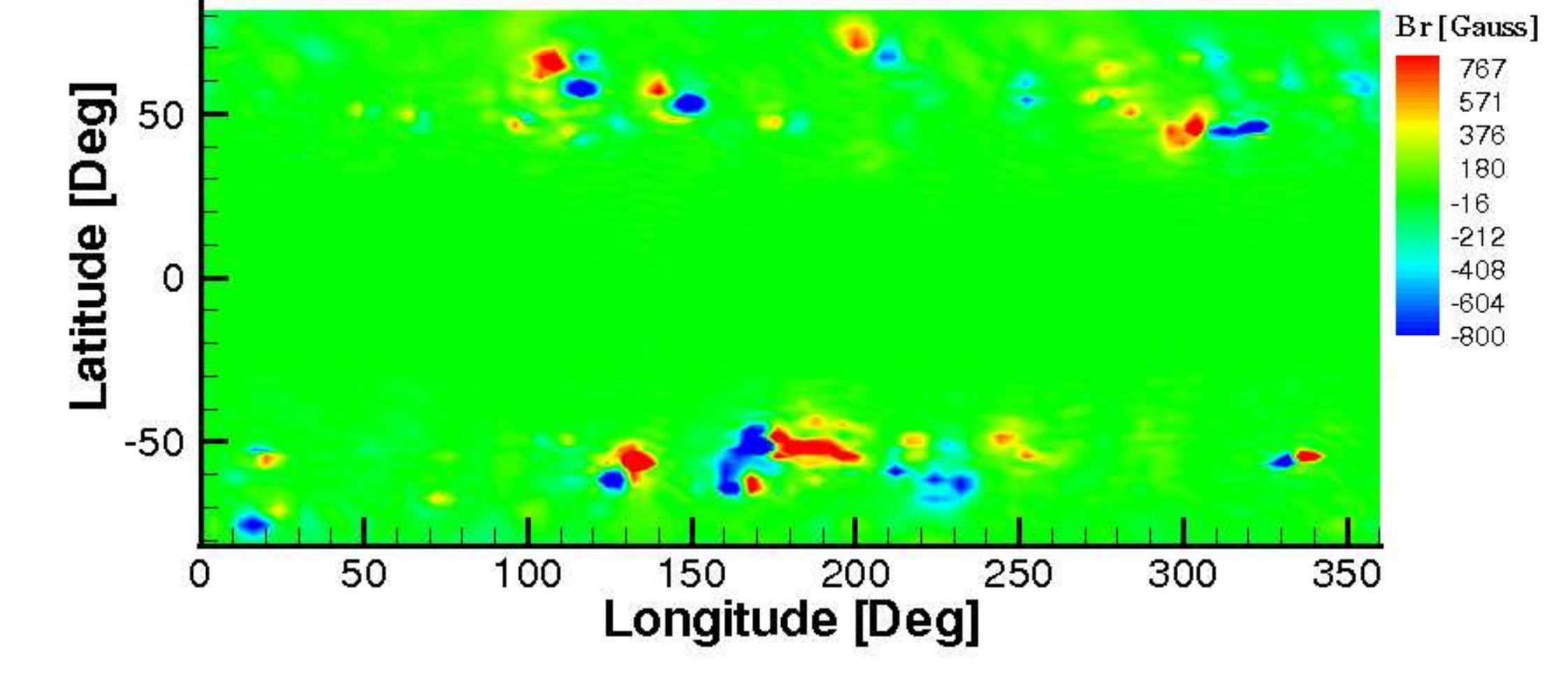}
\caption{Fiducial stellar surface magnetic maps for low (30 degrees; as observed on the Sun) and high (60 degrees)
  latitude magnetic active regions. }
\label{fig:2a}
\end{figure*}

We perform three-dimensional stellar wind simulations for surface magnetic
maps of solar active regions observed near solar maximum (CR 1958)
with an underlying dipolar component, using the solar rotation period
($\sim 25$ days).  The CR 1958 magnetogram is first artificially
dismembered so as to separate out the large-scale field and the
small-scale spots. Given that the spots have much stronger field than the
background, we distinguish between these two components by imposing a
threshold of 10 G and discard any weaker field than that.  This provides
us with a magnetogram of the magnetic active regions only.  On the
other hand we take dipolar magnetograms of different strengths.      
We can then manipulate these two different types of magnetograms and
re-assemble the spots and dipolar components in arbitrary ways.  Within this topological configuration, we
explore three parameters: the latitudinal position of the
spots, and the absolute and relative strengths of the dipolar field and
the active regions.  Representative magnetograms for the cases of the
observed solar spot distribution (30 degrees latitude) and one with the
spots shifted to high latitudes (60 degrees) are illustrated in Fig.~\ref{fig:2a}.

Our grid of fiducial magnetograms consists of 10 and 20 Gauss dipolar fields with spot
strengths ranging from the original sunspot magnetic fields up to four times
the observed one. This provides six combinations of different relative
strengths in addition to the two pure dipoles. Furthermore, we explore four different
topologies: the pure dipole and the dipole with
spots rings located at 30 (as observed), 60, and 80 degrees of latitude, with an additional case at 50 degrees for the case of a 20~G dipole and observed spot field strengths.  
The spots extracted from the CR 1958 magnetogram do not have a zero net flux.  The origin of this probably lies in the known north-south asymmetry of sunspot activity \citep[e.g.][]{Hathaway:10,Mordvinov.Yazev:13} coupled with 
the limited sensitivity and resolution of the observations and the procedure of disentangling the spots from the large-scale dipolar field.  The imbalance amounts to a few percent, with the opposite sign as that of the underlying dipolar field.  It is possible that this difference in the net flux of each hemisphere given by the residue of the small scale
activity could influence the wind model.
As a quality check, we therefore also compute magnetograms in which the polarity
of the spots is inverted in order to quantify any systematic effects due to the small-scale flux imbalance.
Moreover, we force the dipolar component to remain constant when
including the spots in order to rule out any effects due to the
change in the dipolar flux introduced by the active regions.

\subsection{Mass and Angular Momentum Loss equations}

From the three-dimensional model solutions we extract the wind density and speed over the
Alfv\'en surface and at the stellar surface. The angular momentum loss
rate given by Weber and Davis (1967) (Equation~\ref{WD}) is an
approximation that assumes spherical symmetry. In order to allow for a
more general asymmetrical magnetic and wind morphology, we compute
the mass and angular momentum loss rates at each point of the Alfv\'en
Surface.

The mass loss is straightforward an computed locally by
\begin{equation}
dM/dt = \rho \,  (\mathbf{u} \cdot \mathbf{dA})
\end{equation}
Here, $\rho$ is the local density,
$\mathbf{u}$ is the local velocity of the plasma, and $\mathbf{dA}$
refers to the surface element on the Alfv\'en Surface.

The angular momentum loss is potentially more complicated to compute.  
\citet{Mestel99} considered charged particles escaping through 
a magnetic field.  A flow under such conditions and subject to Lorentz forces leads 
to a  $z$-component of the angular momentum loss that, computed at the
Alfv\'en surface, is given by (Mestel 1999),
\begin{equation}
dJ/dt =  (\mathbf{r} \times \mathbf{dA} )(P +
B^2/8 \pi) +  \Omega \rho R^2 \sin^2{\theta}(\mathbf{u} \cdot \mathbf{dA}),
\end{equation}
where $R$ is the radial component in spherical coordinates and $P$ is the wind gas pressure.
The first term is the net magnetic transport that should occur about the $z$-axis, and was also highlighted by \citet[][see their Appendix]{Vidotto14b}.  However, 
 we have verified that this term is negligible in our model solutions and angular momentum loss equation then
reduces to the effective corotation term
\begin{equation}
dJ/dt = \Omega \, \rho \, R^2 \sin^2{\theta} \, (\mathbf{u} \cdot  \mathbf{dA}),
\end{equation}
where $dJ$ is the component of the angular momentum change in the
direction of the rotation axis, and is the only one contributing to angular momentum loss.
\noindent We also calculate the amount of open magnetic flux through
a spherical surface outside of the Alfv\'en surface for each case,
and map the three-dimensional Alfv\'en surface itself.

\section{RESULTS}
\label{sec:Results}

\begin{figure*}
\center
\includegraphics[width = 4.5in]{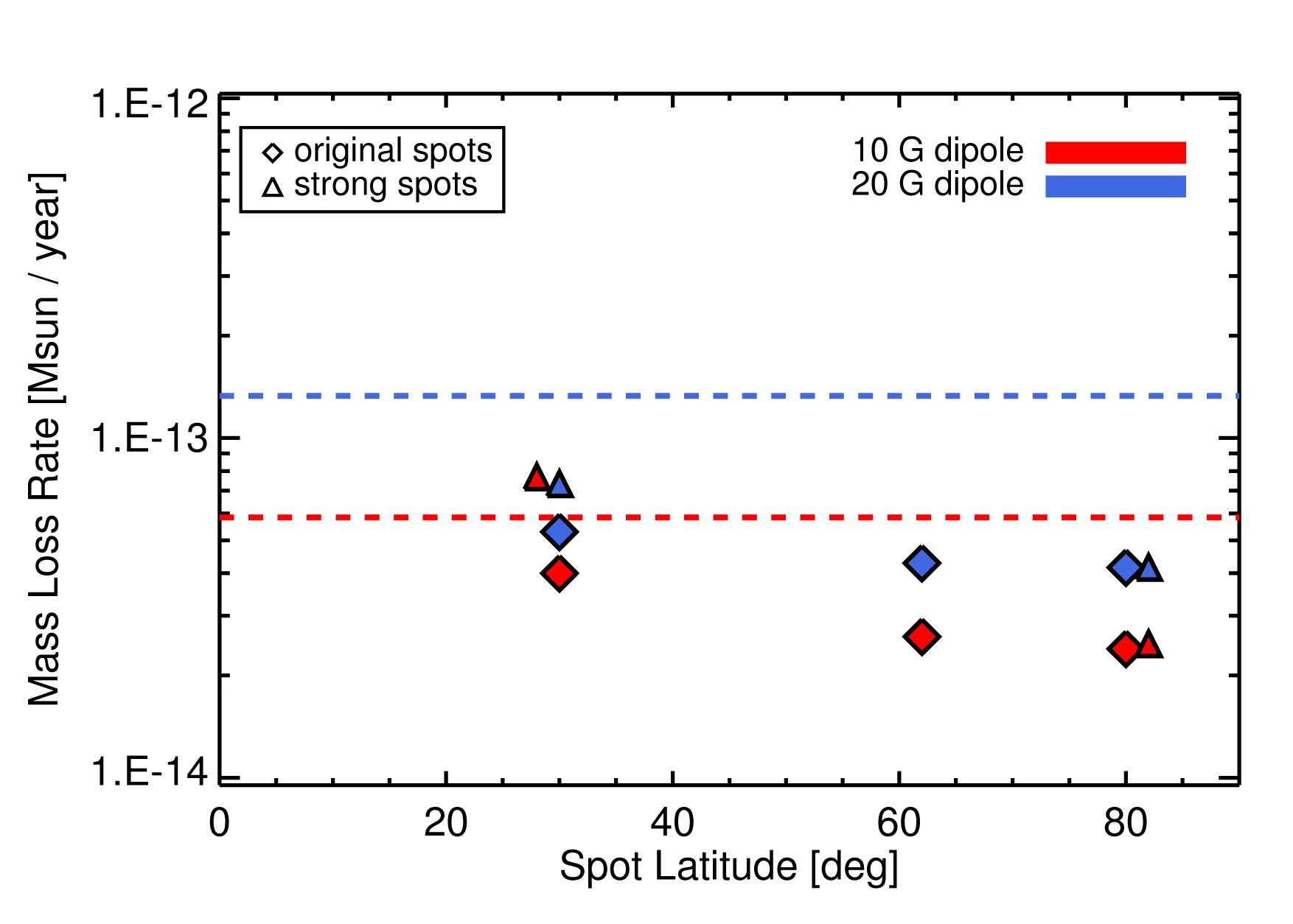}\\
\includegraphics[width = 4.5in]{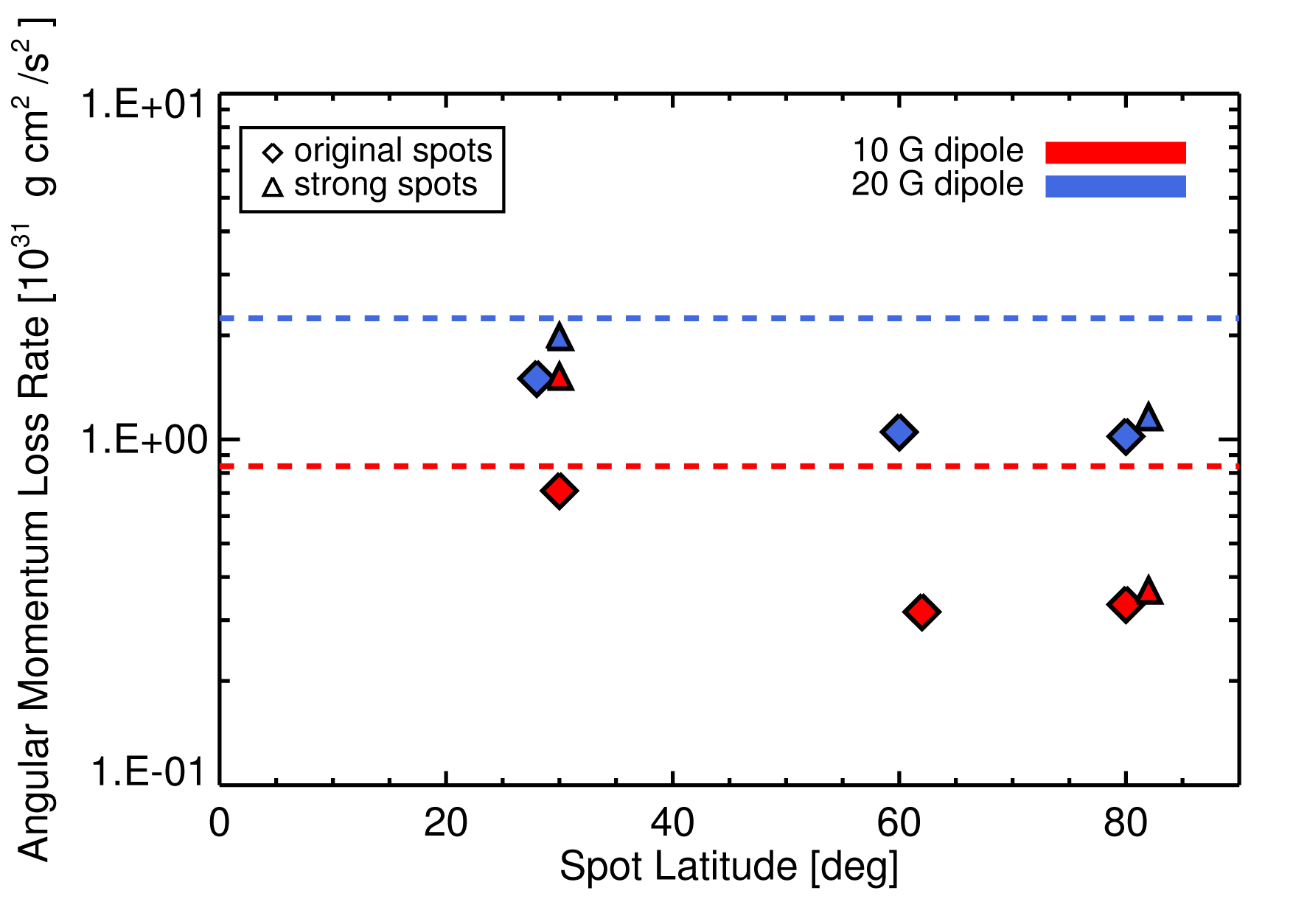}\\
\includegraphics[width = 4.5in]{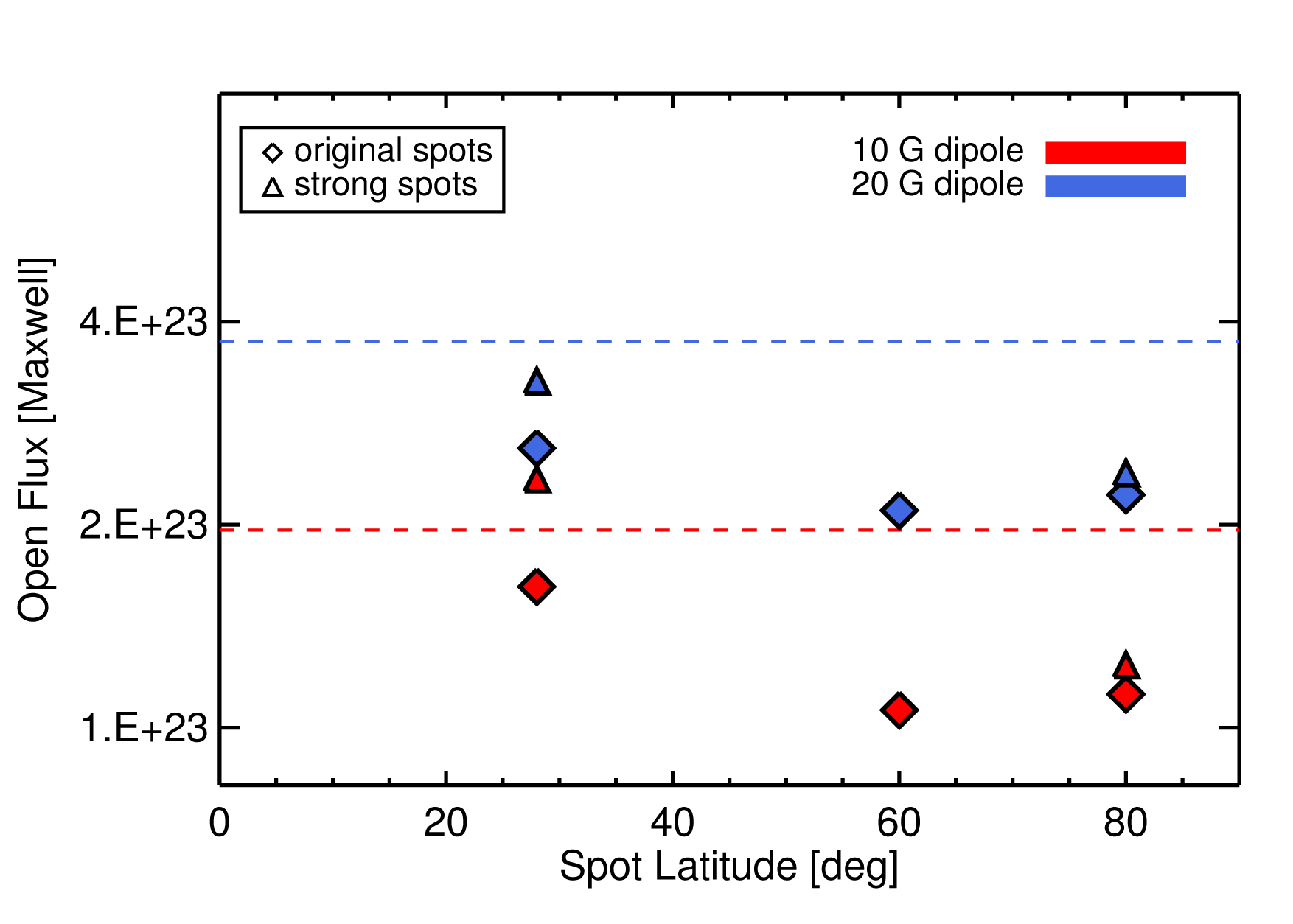}
\caption{Mass (top plot) and angular momentum (middle plot) loss
  rates, and amount of open flux (bottom plot) for different topologies as a function of the latitude of their magnetic active regions (x-axis) with an underlying dipolar field of 10~G (red) and 20~G (blue). 
}
\label{fig:1}
\end{figure*}

The mass and angular momentum loss rates for the grid of magnetograms
are listed in Table~1.  The central cases (observed and strong spots
with their original polarity at
different latitudes for the two dipolar field strengths) are illustrated in
Fig~\ref{fig:1}.  The rest of the cases are just quality checks and
are left out of the plot for clarity.  They all lie very close to
the results shown and the trend is independent of these particularities.  Dipole results are shown as
dashed horizontal lines to serve as a reference.  Differences among
cases are expected, and observed, due to the differences in the resulting magnetic flux and
dipolar component. In the real Sun, however, the total flux balances, so this residual flux is simply an artifact of the model and should not be having a significant effect.  Here, two effects are
entangled: the small-scale activity itself and the change of the
large-scale structure
due to cancellations (or additions) introduced by the net effect of the small-scale structures.
From Fig~\ref{fig:1} it
is clear that, for all the cases, as spots move up in latitude the
mass and angular momentum losses are reduced.  The drop is more pronounced 
when crossing a mid-latitude between 30 and 50 degrees.  The largest effect is
seen for the
scenario in which the strong spots have their original fields and are placed on the
hemisphere with opposite polarity to the one of their residual flux
(triangles in Fig~\ref{fig:1}).
This points to the net residual spot field as an additional agent acting towards a
 reduction of mass and angular momentum losses as spots are moved up in latitude.  
However, even in the opposite case (spots with inverted polarity),
where the dipolar change due to the spots should work towards a compensation of
the small-scale field effect, the overall qualitative behavior is the
same (but slightly
smaller, as expected), indicating that in these cases small-scale structure dominates
over these differences.

Low-latitude active regions lead to a more diverse set of results than
high-latitude ones.  For most cases, there is a decrease in mass loss
and yet, sometimes this results in an increase of the angular momentum
loss. In principle one should expect both quantities to increase and decrease
together, since the angular momentum is getting lost through the mass
loss.  However, the relative orientation of the winds and the rotation
axis can affect the angular momentum loss 
dramatically but not the mass loss rates.  That
these two quantities do not systematically change together when including low
latitude spots points towards a change in the orientation or topology of the
Alfv\'en surface.  In two
of the cases (strong spots on a weak dipolar field with both observed,
triangles in Fig~\ref{fig:1}, and inverted spot polarity, table~\ref{table:aml}) 
there is an increase in mass loss rates when magnetic
active regions are introduced at low latitudes.  
In contrast, high-latitude magnetic spots result in a
reduction of both mass and angular momentum loss rates for all cases.
It is worth pointing out that, as shown in Fig~\ref{fig:1}, for high latitudes, original (diamonds) and strong
spots (triangles) solutions result in almost the same rates.  This
indicates that topology dominates over total magnetic flux in this
regime. 

The lower panel of Fig~\ref{fig:1} shows the change in open magnetic
flux for the same cases. It is interesting to notice that this trend
matches very well the one for mass and angular momentum loss rates
(middle and top panels of the same Fig).  

The general qualitative behavior can, then, be said to depend largely on the
latitude of the spots and can be illustrated by two limiting cases:
low-latitude and high-latitude active regions (see Figure \ref{fig:2a}).

\begin{figure*}
\includegraphics[width = \textwidth]{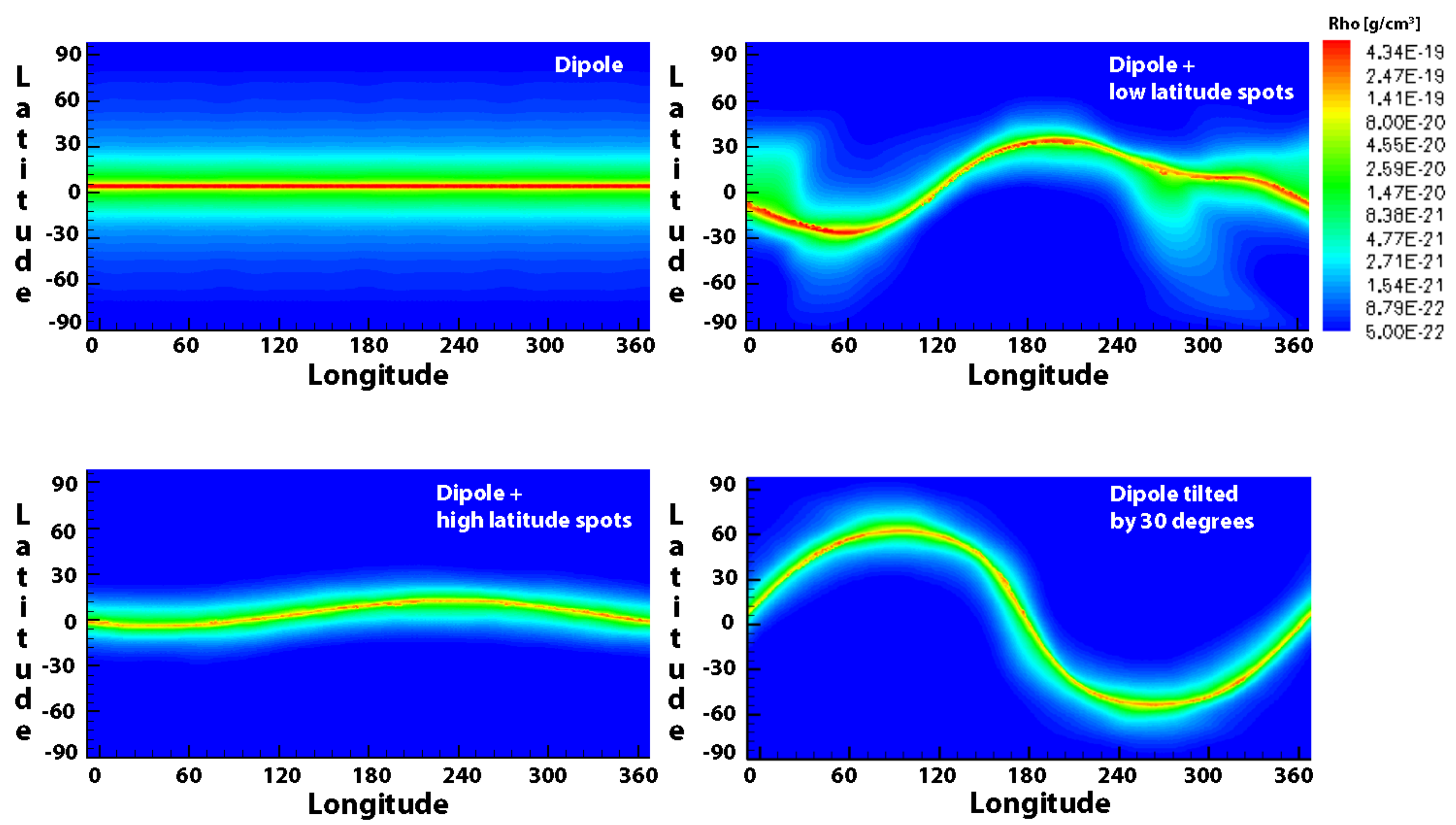}
\caption{Density at the Alfv\'en surface as a function of latitude (x-axis) and
longitude (y-axis) for four of the models that illustrate key aspects
of the results: a pure 10~G dipole; the same dipole with low-latitude
spots of observed strength; the same dipole with high-latitude
spots of observed strength; and a pure 10~G dipole tilted by 30 degrees.}
\label{fig:3}
\end{figure*}

\begin{figure*}
\includegraphics[width = \textwidth]{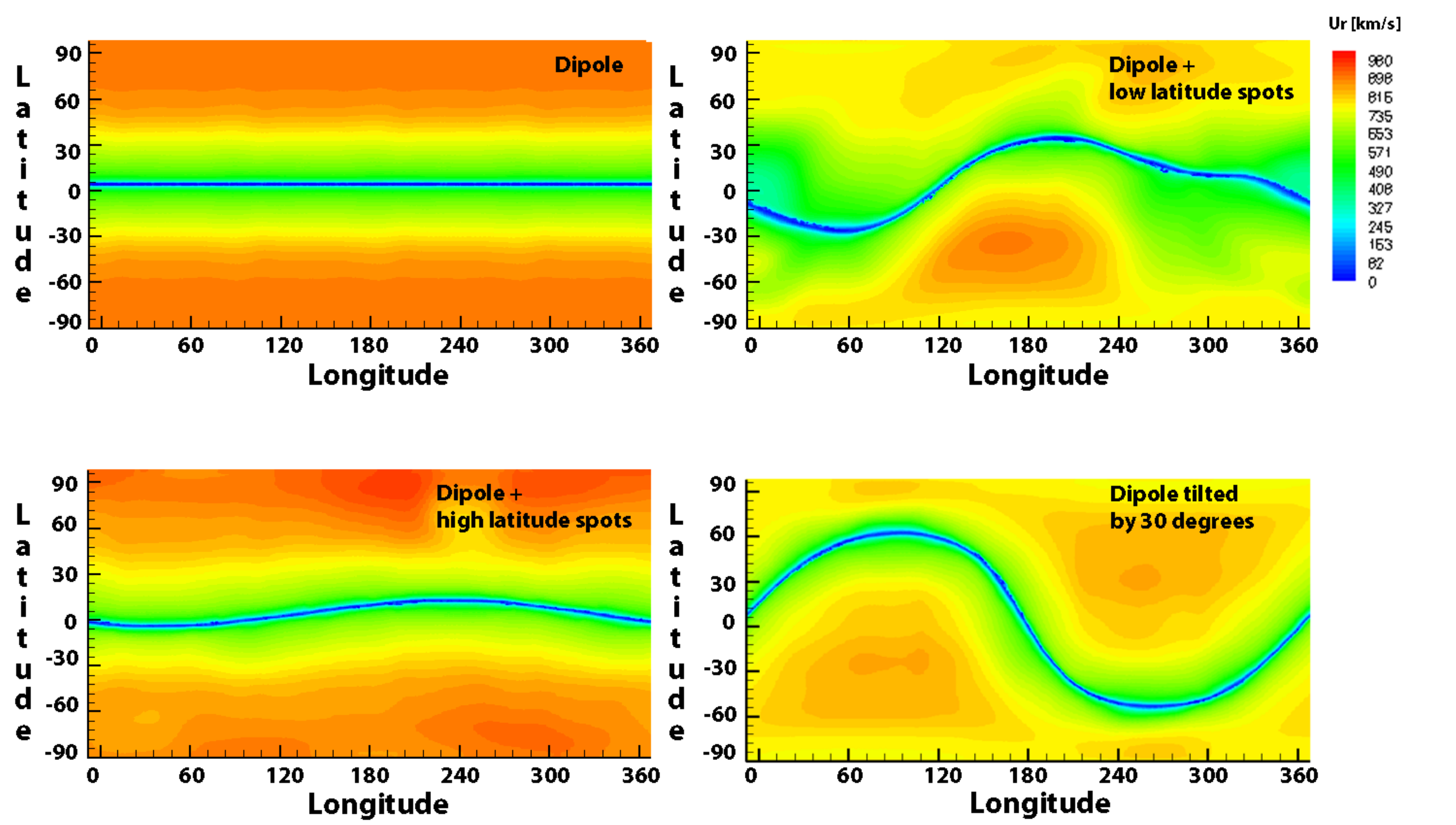} 
\caption{Radial component of the wind Speed at the Alfv\'en surface as a function of latitude (x-axis) and
longitude (y-axis) for the same model cases illustrated in Figure~\ref{fig:3}.}
\label{fig:4}
\end{figure*}

It is useful to visualize the relevant model variables at the Alfv\'en
surface. The two-dimensional projections in latitude and longitude of the density and wind speed
at the Alfv\'en surface are illustrated in Figures \ref{fig:3} and
\ref{fig:4}.   The cases shown are the pure dipole (with a field
strength of 10~Gauss), the same dipole with
low-latitude spots, the same dipole with original field strength high-latitude spots, and a tilted
pure dipole of the same strength
to serve as a reference.  

\begin{figure*}
\includegraphics[width = \textwidth]{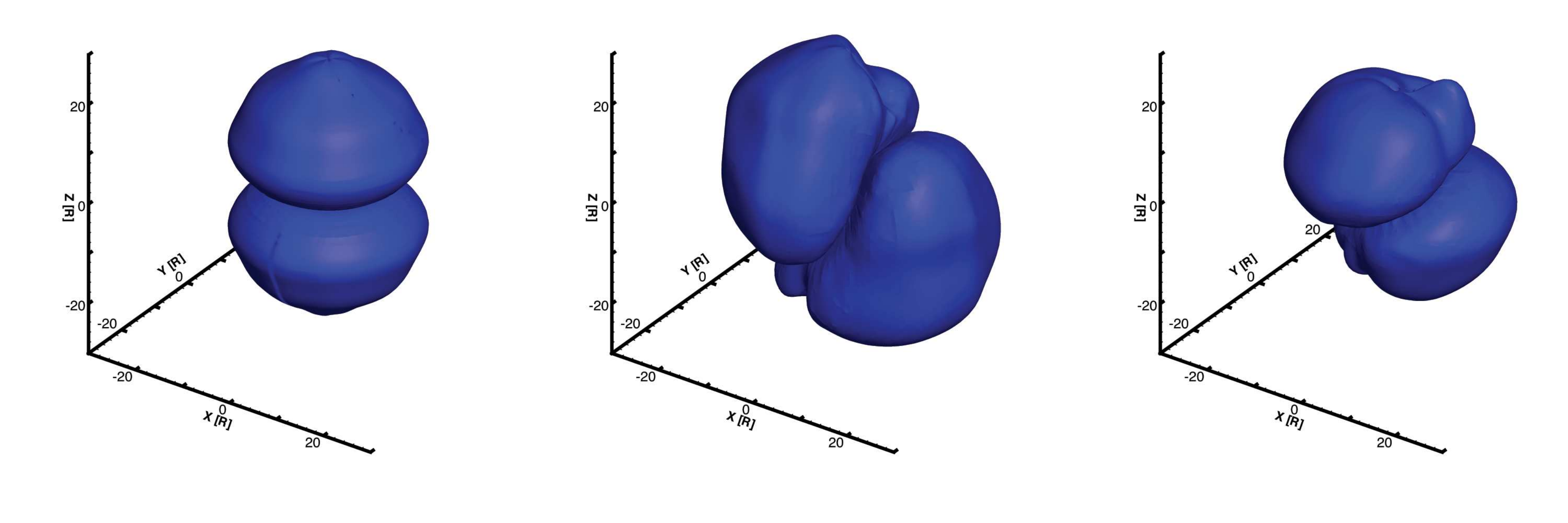}
  \caption{Three dimensional Alfv\'en surface in units of stellar radii for a dipole (left),
  low-latitude magnetic active regions (center), and high-latitude
  magnetic active regions.}
\label{fig:5}
\end{figure*}

When introducing low-latitude spots, the
distributions of the density and wind speeds mimic that of the tilted
dipole case.  This is essentially because of the perturbation to the large-scale field arising from the active regions.  As these active regions are shifted towards
higher latitudes, one recovers the solution of a scaled down
dipole. The same behavior can be confirmed from the shape and orientation of the
Alfv\'en surface for each case. As Fig~\ref{fig:5} shows, the
Alfv\'en surface gets tilted when low latitude spots are introduced, and the tilt angle is decreased as spots are moved to higher latitudes. The tilt angle also 
depends on the relative strength of the spots and the underlying
dipole. 

In Fig~\ref{fig:6}, we show the mass and angular momentum loss rates as a
function of latitude for the 10~G dipole, and
the two limiting cases of low-latitude (30 degrees) and high latitude (80 degrees) spots in one of the
scenarios of our grid: that with an underlying 10 Gauss dipole and in which the 
spots are four times stronger than solar and are situated in their observed hemisphere where polarity is opposite to their residual polarity (the qualitative effect is the same for the
whole sample). The solid line corresponds to measurements at
the Alfv\'en surface and the dotted line at the stellar surface. 
Both mass and angular momentum losses decrease when the active regions
are shifted towards higher latitudes. 

Mass loss is very strongly peaked at the equator of the Alfv\'en
surface for the dipole and for the case of low-latitude spots.  As is then expected, angular momentum loss also peaks strongly at the
equator but with much more breadth in latitude.
  However, it is clear when looking more closely at the star itself that this mass
is primarily coming from mid-latitudes and becomes equatorial as it
approaches the Alfv\'en surface and is bent around the closed ``dead zone'' dipole loops of the equatorial field.   This implies quite a striking result:  the torque experienced by the star is not exerted over the whole of its surface, but 
will be confined to narrow mid-latitude bands.  It is also worth
noting that these bands are
where the larger mass loss differences can be found when comparing low and
high latitude spot topologies, and correspond to the
limiting latitude between the closed and open field line areas of the
dipolar wind solution.


\section{DISCUSSION}
\label{sec:Discussion}

The main aim of this study is to investigate the influence of small-scale magnetic field in the form of sunspot-like starspots on the mass loss and angular momentum loss of stars, and how these loss rates vary over the stellar surface.  While we have presented models based on a specific active Sun magnetogram, the findings of the modeling presented here should be quite general to stars with active regions not too dissimilar to those of the Sun, and with underlying low-order large-scale fields.  This work can be considered an extension to the idealized study of the influence of spots and their latitudes on stellar winds by \citet{Cohen09}, and to the exploration of winds driven by purely dipolar fields by \citet{Cohen.Drake:14}.

\subsection{General results}

We find a bi-modal regime for the wind solution of solar-like stars
that depends on the latitudinal location of
their active regions.  The pure dipole solution has a ``dead zone" at low latitudes in which the wind is essentially trapped and does not escape. 
While low-latitude spots within the dead zone mimic a tilted dipole
solution, active regions at higher
latitudes result in a decrease of both mass and angular momentum loss rates.
 The limit between the two regimes is the transition latitude from closed to open
field lines of the equivalent dipolar solution (i.e.\ the one for the same surface
magnetic map without small scale active regions).
 The acting mechanisms that lead to these two different qualitative
behaviors are illustrated in Fig~\ref{fig:7} and
are the following: a competing non-aligned dipole, and closing of otherwise open
field lines that cuts off regions that would otherwise participate in wind loss.  

The effect of introducing active regions within the dead-zone is
mainly a tilt of the dipolar solution.  Any realistic distribution of
spots will have a residual dipolar component because of both azimuthal asymmetry in the general spot locations, and the tilt  with respect to the equator of the spot pairs with opposite polarities as they emerge at the solar (or stellar) surface (Joy's Law).  In the case of a
low-latitude ring of active regions with significant azimuthal asymmetry, this component is almost
perpendicular to the original dipolar one and serves
as a competing dipole. 
When added to the original underlying dipole, this will result in
a tilt to the magnetic axis, and will be larger for larger ratios between spot and dipole field strengths.  Because angular momentum loss is sensitive to the latitude at which the wind escapes the Alfv\'en surface, a
change in angular momentum loss is expected due to the misalignment of
the rotation and magnetic axes.  

Spots at low latitudes usually do not
have the capacity of closing open field lines since they reside in a region of closed field lines.
However, for strong spot fields, spots might be able to couple to a few of
the lower latitude open field lines leading to a moderate reduction
in mass loss rate.  
In some cases, if the spots have much stronger fields than that of the
underlying dipole, there could be a significant change in the 
mass loss due to a change in the large scale background field.  
This is the case for strong spots introduced on the weakest dipole (see Fig~\ref{fig:1}).
The behavior persists for inverted spot polarities, however, this is not
surprising since the spots are located near the equator and, therefore, the
difference in the background field of the ones situated in the northern and southern
hemisphere is small.  As expected, the effect
disappears when keeping the dipolar (large scale field) component constant.

In contrast, it is clear from the open flux behavior (bottom panel of Fig~\ref{fig:1}),
that when the magnetic active regions are placed
at higher latitudes they close otherwise open field lines (as sketched
in Fig~\ref{fig:7}) leading to a
reduction in mass and angular momentum loss rates.  Only open field
lines are able to carry away mass, and most of this mass comes from the
lowest latitudes within the open field regions (see
Fig~\ref{fig:6}). This is not surprising since it is well known that
fast polar winds are lower in density and provide a less efficient
mass and angular momentum loss mechanism.
Fig~\ref{fig:6} shows that for high-latitude spots, the largest
reduction of mass loss rates, as compared to the dipolar case, is at mid-latitudes, at 
the beginning of the region of open field lines from where most of the
mass loss originates.  This supports our finding that it is the closing
of open field lines that leads to a reduction in mass and angular momentum loss rates.   
The efficiency of
this mechanism increases with the strength of the magnetic field of both the dipole
and the spots.  The reason is that the stronger the field the
larger the distance over which two opposite polarity footpoints can interact.   

The effect of spots in closing down regions of open field echoes the
findings of \citet{Cohen09} based on a synthetic distribution of ideal
spots.  However, we note that some aspects of the results presented
here are different to the predictions of  \citet{Cohen09}.  In
particular, the trend of increasing mass and angular momentum loss
with increasing  spot latitude for very strong 1kG synthetic spots
found in that study  is opposite to the trend we find here.
We have re-run our simulations using the same method as
\cite{Cohen09}.  When comparing the results the old model shows an increase in
the mass-loss rate from the base dipole case. This is the result of
the artificial scaling of the base density with the field line
expansion factor in the old model, which was designed to provide a
solution that compares better with observations of the solar wind. It
is scaled to have higher density at the base of flux tubes that are
associated with larger expansion, which occurs when small-scale strong
field is introduced. As a result, the more spots we add, we
artificially boost the overall mass that is provided to the system and
as a result, the total mass-loss rate is higher with more spots than
for the case of no spots. In the new model, the base density is
uniform and the density modulations are generated self-consistently as
a result of the coronal heating implemented in the model. Therefore,
the mass source for all cases is the same and the cases with spots
show reduction of the mass-loss rate compared to the dipole case.  In
\cite{Cohen09}, the point corresponding to strong spots at 30 degrees
(Figure 6 top panel) uniquely increases the source density due to the
fact that the spots are right at the edge of the large helmet
streamers, and the overall mass-loss is highest (this is not reflected
in the angular momentum loss rate).  Except for this point, the overall
trends in the old and new models are consistent although difficult to
compare due to the different latitudes explored.

\subsection{Relevance for different activity levels and stellar cycles}

Both observations and theoretical models suggest that high-latitude
magnetic features should be
expected for rapidly rotating, magnetically active stars
\citep{Schuessler.Solanki:92,Schrijver.Title:01}.  High-resolution X-ray spectroscopy of the young, rapidly rotating K0~V star AB~Doradus has also revealed evidence for dominant polar coronal emission \citep{Drake.etal:15}.
 This quite fundamental difference compared with the solar magnetic field raises the obvious question of how the mass and angular momentum losses 
might be influenced by the magnetic topology and whether this could be relevant for stars in the saturated activity regime \citep[e.g.][]{Vilhu:84,Vilhu.Walter:87,Wright.etal:11}.  The spin-down of saturated activity stars appears to be generally slower than expected based on extrapolation of the \citet{Skumanich72} law, and this has historically been interpreted in terms of a fairly arbitrary saturation in the efficiency of angular momentum loss \citep[e.g.][]{Kawaler88,Chaboyer95a}.
However, young stars in open clusters are observed to have a bimodal distribution of rotation rates in which the rapid rotation of some of them appears to persist longer than expected without any efficient mechanism to curb it \citep[e.g.][]{Stauffer87,Soderblom93,Meibom11}.

In the above context, \cite{Solanki97}, \cite{Holzwarth05a}, and \cite{Holzwarth05b} have investigated the relevance of the 
latitudinal distribution of magnetic flux on the rotational evolution of
cool stars. They found that the assumption of a spherically-symmetric wind is generally not adequate to describe mass and angular momentum loss, and that high latitude magnetic flux can act to quell the angular momentum loss.  

In the light of the recent finding by \citet{Garraffo.etal:15},
  that increasing the magnetic complexity of the large scale magnetic fields on
  stellar surfaces can lead to an effective suppression of mass and angular
  momentum loss by 2 to 3 orders of magnitude, we find that the
  contribution of the magnetic active regions is more of a second order effect, but one 
  that can act to further reduce the efficiency of the braking, by
  factors of a few.  A reduction in both mass and angular momentum losses 
can be expected from high-latitude active regions closing
otherwise open field lines and, therefore, shutting
down some of the channels for magnetic braking.  
 While the effect for the magnetic maps used here are not dramatic at 50--100\%\ or so in terms of angular momentum loss rates, at early times these differences translate to similar differences in expected rotation rate as a function of age.   Due to this type of sensitivity, it will be very difficult for any {\it ab initio} models of stellar spin-down based on wind modeling to correctly predict the spin down rates of stars to useful precision without being calibrated by observations.  It is possible that long-lived stochastic differences in magnetic topologies for otherwise fairly similar stars could contribute at least some of the large dispersion in the correlation
between average magnetic field strength and rotation period observed
by \citet{Vidotto14a}.  Since the number, size and location of magnetic active regions is 
correlated with stellar age and rotation period, we can
expect an age differentiated effect, with a larger scatter for younger, more rapidly rotating stars.   

Based on our experience with the sun, one should expect the latitude of active
regions to change also during stellar
cycles. If during the magnetic cycle of a star its active regions cross
the limiting latitude between open and closed field line areas (unlike in the sun where the spots move exclusively within the dead-zone), their mass and angular
momentum loss rates could undergo a sudden change of by a factor of a few.  Magnetic modulation of angular
momentum loss could then be expected
and caution should be taken when assuming single snapshot measurements of
stellar mass loss rates to be representative of average rates.
The magnitude of the effect we find here is still an order of
magnitude too small to explain the recent \cite{Wood14} deduction of a
weak wind from the active star $\pi^1$~UMa (and the inference of such
on $\xi$~Boo~A), although we note that the limited case of solar-like
magnetic field topologies we treat here are unlikely to be appropriate
for all very active stars.   ZDI stellar observations suggest
that the average magnetic field strength changes over time as $<|B_v|
\propto t^{-0.655 \pm 0.045} > $ \citep{Vidotto14a}, supporting a linear-type dynamo of
the large scale field. However, this change is expected to be much
slower than the stellar cycle. In this work we kept the dipolar
component constant in order to disentangle both effects and study the
mass and angular momentum loss changes due only to the different
latitude of the active regions.   It is worth pointing out that there could
be an interplay between the spot migration and the underlying dipolar
strength, as is the case for our sun.

For the case of our sun, \cite{Cohen11} finds a nearly constant mass loss rate scattered around
$2 \times 10^{-14} \,M_{\sun} \, yr^{-1} $, independent of the
magnetic cycle or flux variations. Furthermore, the solar and heliospheric open magnetic flux is
  modulated by only a factor of 2--4 over the solar cycle compared to
  the total magnetic flux, which is modulated by more than a factor of
  10  \citep[e.g.][ with references therein]{Wang.etal:00,Owens.etal:11}. The
  addition of open flux during enhanced solar activity is attributed
  to coronal mass ejections, which result from
  interchange magnetic reconnection processes
  \citep{Crooker.etal:02}. The loss of open flux is attributed to flux
  cancellation due to footpoint motions in the photosphere and to
  differential rotation. Both of these processes are dynamic, and thus
  are not captured by our steady-state solutions. Nevertheless, our
  simulation for the unaltered solar magnetogram (10~G dipole with
  spots at 30 degrees) recovers rather well the observed solar open flux of
  $0.75-1.25\cdot 10^{23}\;Mx$.  Solar mass loss and open flux
  observations are also consistent
  with our results since the active regions of the sun remain within
  the dead-zone during the whole cycle and, therefore, are not efficient at closing open field lines.  In contrast, our
  results and the open flux solar observations seem to be
 inconsistent with \citet{Reville.etal:15}, who find a difference of
  an order of magnitude on the open flux between solar maximum and
  solar minimum leading to a different torque acting on the Sun
  at these two stages.

 \citet{Catala07}, \citet{Donati08}, and \citet{ Fares09} provided magnetic maps for $\tau$
Boo, which show a dominant poloidal field of $5 G$ with the probable
presence of a small toroidal component. They further note that active regions
being located at different latitudes during different
seasons is likely.
\cite{Vidotto11, Vidotto12} used these magnetograms in combination with a Parker-type wind and found a factor of two variation in angular
momentum loss rates over the magnetic cycle. Furthermore, they find that the angular
momentum loss rate decreases for magnetic maps with flux concentrated at higher
latitude.  Although
low-order field magnetic maps where used in their simulations, the results are 
in accord with expectations from the simulations presented here, as shown in \ref{fig:6}. In our case we argue that the reason for this 
reduction is the closing of otherwise open field lines.

\subsection{Latitude of wind torque}

We have shown in Fig~\ref{fig:6} that mass loss occurs predominately
at mid-latitudes, and therefore angular momentum torque will be
similarly exerted at mid-latitudes.  Could this preferential latitude
for the exertion of spin-down torque have observable consequences on
the outer layers of a star? The energetics of the solar wind are,
however,  orders of magnitude smaller than even second order effects
such as cyclic variations in the meridional flow on the Sun.
Assuming, very roughly, that a few percent of the solar mass might be
involved in such flows, and that variations in flow velocities are of
the order of a few m\,s$^{-1}$
\citep{Komm:94,Ulrich.Boyden:05,Gizon.Rempel:08,Komm.etal:11}, the
energy involved in the variations is about $10^{36}$~erg~s$^{-1}$,
compared with a few $10^{27}$~erg~s$^{-1}$ in solar wind kinetic
energy.   Similarly, the maximum angular momentum loss rates we obtain
in our present simulations are about $10^{38}$~g~cm$^{2}$~s$^{-1}$
per year, while about 1\% of the solar mass rotating close to $1R_\odot$ at a few m~s$^{-1}$ is about $10^{44}$~g~cm$^{2}$~s$^{-1}$.  The situation for more active stars, even if they have more active winds, will be similar.  It then appears that the exact latitude of emergence of the wind from the star is not important and the star is rotationally braked essentially as a solid body.


\section{SUMMARY and CONCLUSIONS}
\label{sec:Conclusions}

We have investigated the role of active regions on the mass loss and spin down
rates of solar-like stars. We find a bimodal regime regulated by the latitude of the magnetic
active regions and by the residual dipolar component of their combined fields.  
Low-latitude active regions result in a tilt of the magnetic axis and
Alfv\'en surface determined by the ratio of spot field strength to dipolar
field strength, while high-latitude active regions lead to a
systematic suppression of mass and angular momentum losses with an 
efficiency that depends on the absolute values of the dipole and spot
magnetic field strengths.  The sensitivity of mass and angular momentum loss rates to the details of magnetic topology present a problem for {\it ab initio} models of stellar spin-down correctly predicting the spin down rates of stars to useful precision.

Although the Alfv\'en surface is the natural place to calculate the
angular momentum loss, the mass being lost has its origin at the stellar
surface.  We have explored these foot points and
have found
that the mass carried away through the winds originates primarily in a
narrow band of mid latitudes at the limit of the dead zone of close equatorial field.  
This means that the torque experienced by the star
is being applied at preferential latitudes, although this is unlikely to be 
important for the structure or dynamics of the outer stellar layers.


\acknowledgments

\begin{figure*}
\includegraphics[width=\textwidth]{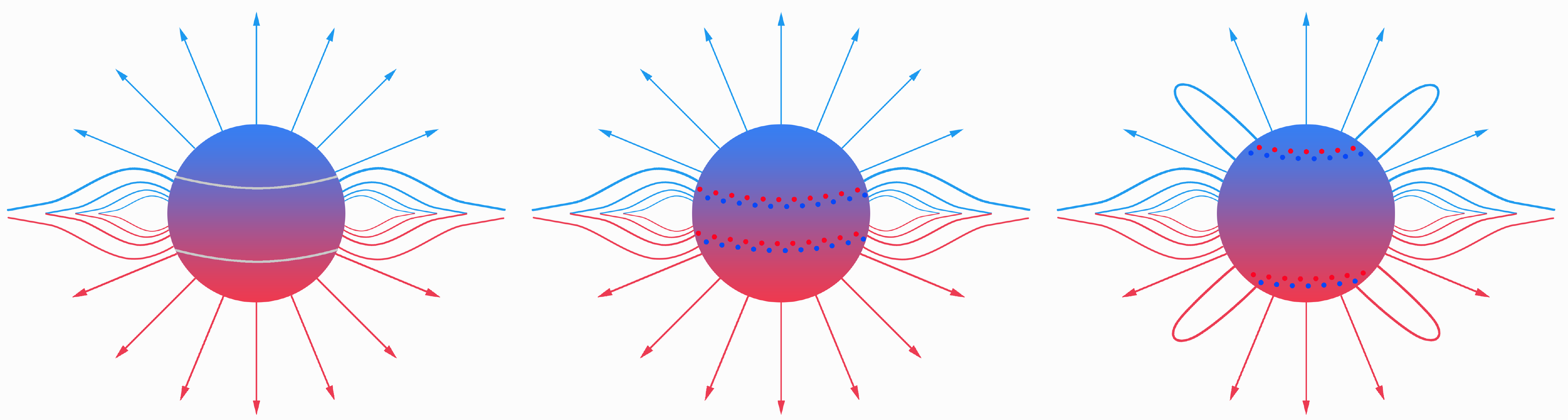}
\caption{Qualitative plot of wind structure for a dipole (left), and the same
  with low-latitude (center) and high-latitude (right) magnetic active
  regions. The limiting latitude between the open and closed field line
  regimes of the dipole is plotted in white (left).}
\label{fig:7}
\end{figure*}

\begin{figure*}
\includegraphics[width = \textwidth]{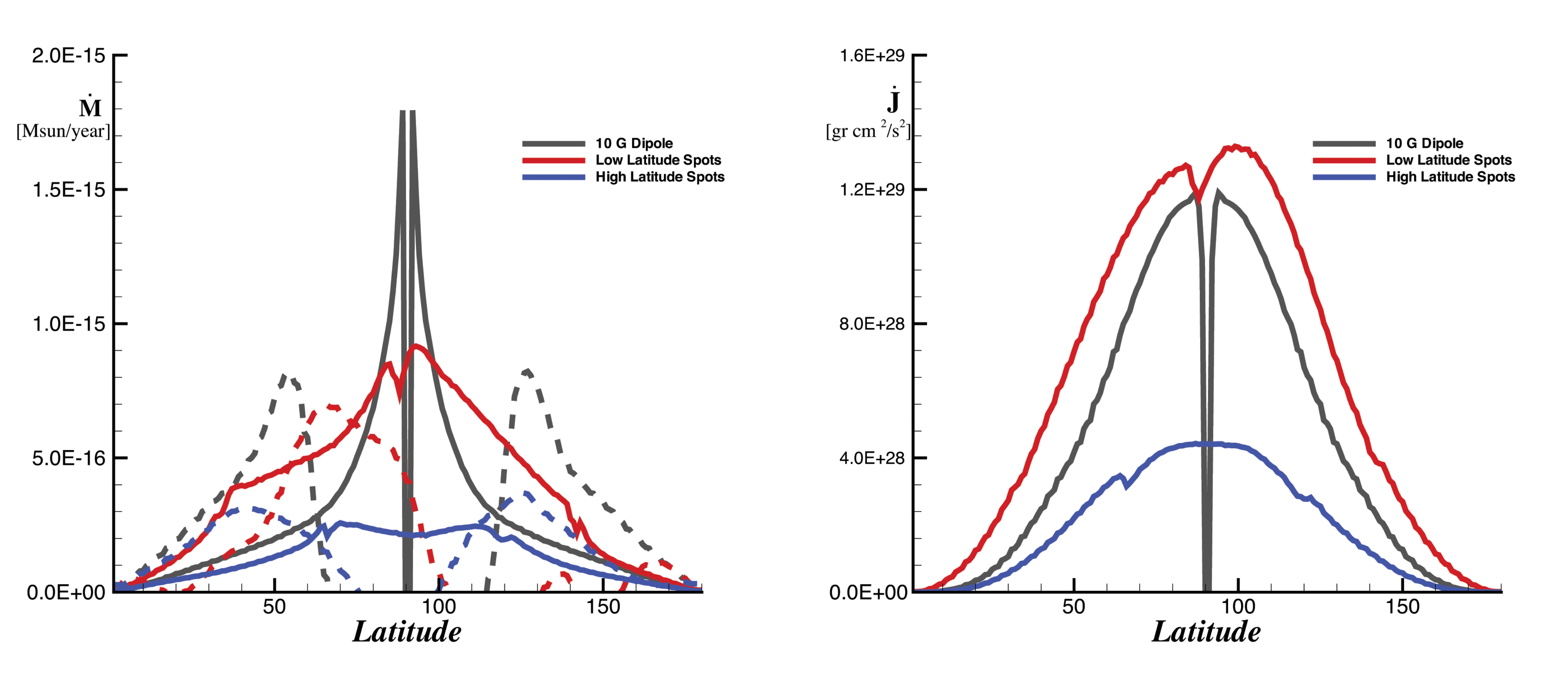}
\caption{Mass and angular momentum losses as a function of latitude, and integrated over longitude, at
  the Alfv\'en surface (solid line) and at the stellar surface (dashed
line) for low-latitude (red) and high-latitude (blue) magnetic active
regions, and for the pure underlying dipole (gray).}
\label{fig:6}
\end{figure*}

We thank Leon Mestel, Jonathan Mestel, Lorenzo Sironi, Steven Cranmer,
and Gaston Giribet for useful discussions. We also thank the anonymous referee
for providing us with constructive comments and suggestions.  We thank Jean-Ren\'e Galarneau for providing Figure~\ref{fig:7}  CG OC were supported by Smithsonian Institution Consortium for Unlocking the Mysteries of the Universe grant ``Lessons from Mars: Are Habitable Atmospheres on Planets around M Dwarfs Viable?'' during the course of this research.   OC was also supported by SI CGPS grant ``Can Exoplanets Around Red Dwarfs Maintain Habitable Atmospheres?''.  JJD was supported by NASA contract NAS8-03060 to the {\it Chandra X-ray Center}, and thanks the Director, B.~Wilkes, and the CXC science team for advice and support.  Numerical simulations were performed on the NASA HEC Pleiades system under award SMD-13-4526.



\begin{thebibliography}{90}
\expandafter\ifx\csname natexlab\endcsname\relax\def\natexlab#1{#1}\fi

\bibitem[{{Allain} {et~al.}(1996){Allain}, {Bouvier}, {Prosser}, {Marschall},
  \& {Laaksonen}}]{Allain96}
{Allain}, S., {Bouvier}, J., {Prosser}, C., {Marschall}, L.~A., \& {Laaksonen},
  B.~D. 1996, \aap, 305, 498

\bibitem[{{Altschuler} \& {Newkirk}(1969)}]{Altschuler67}
{Altschuler}, M.~D., \& {Newkirk}, G. 1969, \solphys, 9, 131

\bibitem[{{Barnes}(2003)}]{Barnes03}
{Barnes}, S.~A. 2003, \apj, 586, 464

\bibitem[{{Bertin} {et~al.}(1995){Bertin}, {Lamers}, {Vidal-Madjar}, {Ferlet},
  \& {Lallement}}]{Bertin95}
{Bertin}, P., {Lamers}, H.~J.~G.~L.~M., {Vidal-Madjar}, A., {Ferlet}, R., \&
  {Lallement}, R. 1995, \aap, 302, 899

\bibitem[{{Bouvier} {et~al.}(2013){Bouvier}, {Matt}, {Mohanty}, {Scholz},
  {Stassun}, \& {Zanni}}]{Bouvier13}
{Bouvier}, J., {Matt}, S.~P., {Mohanty}, S., {Scholz}, A., {Stassun}, K.~G., \&
  {Zanni}, C. 2013, ArXiv e-prints

\bibitem[{{Camenzind}(1990)}]{Camenzind90}
{Camenzind}, M. 1990, in Reviews in Modern Astronomy, Vol.~3, Reviews in Modern
  Astronomy, ed. G.~{Klare}, 234--265

\bibitem[{{Catala} {et~al.}(2007){Catala}, {Donati}, {Shkolnik}, {Bohlender},
  \& {Alecian}}]{Catala07}
{Catala}, C., {Donati}, J.-F., {Shkolnik}, E., {Bohlender}, D., \& {Alecian},
  E. 2007, \mnras, 374, L42

\bibitem[{{Chaboyer} {et~al.}(1995){Chaboyer}, {Demarque}, \&
  {Pinsonneault}}]{Chaboyer95a}
{Chaboyer}, B., {Demarque}, P., \& {Pinsonneault}, M.~H. 1995, \apj, 441, 865

\bibitem[{{Cohen} {et~al.}(1982){Cohen}, {Bieging}, \& {Schwartz}}]{Cohen82}
{Cohen}, M., {Bieging}, J.~H., \& {Schwartz}, P.~R. 1982, \apj, 253, 707

\bibitem[{{Cohen}(2011)}]{Cohen11}
{Cohen}, O. 2011, \mnras, 417, 2592

\bibitem[{{Cohen} \& {Drake}(2014)}]{Cohen.Drake:14}
{Cohen}, O., \& {Drake}, J.~J. 2014, \apj, 783, 55

\bibitem[{{Cohen} {et~al.}(2009){Cohen}, {Drake}, {Kashyap}, \&
  {Gombosi}}]{Cohen09}
{Cohen}, O., {Drake}, J.~J., {Kashyap}, V.~L., \& {Gombosi}, T.~I. 2009, \apj,
  699, 1501

\bibitem[{{Cohen} {et~al.}(2007){Cohen}, {Sokolov}, {Roussev}, {Arge},
  {Manchester}, {Gombosi}, {Frazin}, {Park}, {Butala}, {Kamalabadi}, \&
  {Velli}}]{Cohen07}
{Cohen}, O., {Sokolov}, I.~V., {Roussev}, I.~I., {Arge}, C.~N., {Manchester},
  W.~B., {Gombosi}, T.~I., {Frazin}, R.~A., {Park}, H., {Butala}, M.~D.,
  {Kamalabadi}, F., \& {Velli}, M. 2007, \apjl, 654, L163

\bibitem[{{Cranmer} \& {Saar}(2011)}]{Cranmer11}
{Cranmer}, S.~R., \& {Saar}, S.~H. 2011, \apj, 741, 54

\bibitem[{{Crooker} {et~al.}(2002){Crooker}, {Gosling}, \&
  {Kahler}}]{Crooker.etal:02}
{Crooker}, N.~U., {Gosling}, J.~T., \& {Kahler}, S.~W. 2002, Journal of
  Geophysical Research (Space Physics), 107, 1028

\bibitem[{{Donati}(2003)}]{Donati03}
{Donati}, J.-F. 2003, in EAS Publications Series, Vol.~9, EAS Publications
  Series, ed. J.~{Arnaud} \& N.~{Meunier}, 169

\bibitem[{{Donati} \& {Brown}(1997)}]{Donati97}
{Donati}, J.-F., \& {Brown}, S.~F. 1997, \aap, 326, 1135

\bibitem[{{Donati} \& {Landstreet}(2009{\natexlab{a}})}]{Donati09}
{Donati}, J.-F., \& {Landstreet}, J.~D. 2009{\natexlab{a}}, \araa, 47, 333

\bibitem[{{Donati} \& {Landstreet}(2009{\natexlab{b}})}]{Donati.Landstreet:09}
---. 2009{\natexlab{b}}, \araa, 47, 333

\bibitem[{{Donati} {et~al.}(2008){Donati}, {Moutou}, {Far{\`e}s}, {Bohlender},
  {Catala}, {Deleuil}, {Shkolnik}, {Collier Cameron}, {Jardine}, \&
  {Walker}}]{Donati08}
{Donati}, J.-F., {Moutou}, C., {Far{\`e}s}, R., {Bohlender}, D., {Catala}, C.,
  {Deleuil}, M., {Shkolnik}, E., {Collier Cameron}, A., {Jardine}, M.~M., \&
  {Walker}, G.~A.~H. 2008, \mnras, 385, 1179

\bibitem[{{Drake} {et~al.}(2015){Drake}, {Chung}, {Kashyap}, \&
  {Garcia-Alvarez}}]{Drake.etal:15}
{Drake}, J.~J., {Chung}, S.~M., {Kashyap}, V.~L., \& {Garcia-Alvarez}, D. 2015,
  \apj, 802, 62

\bibitem[{{Fares} {et~al.}(2009){Fares}, {Donati}, {Moutou}, {Bohlender},
  {Catala}, {Deleuil}, {Shkolnik}, {Collier Cameron}, {Jardine}, \&
  {Walker}}]{Fares09}
{Fares}, R., {Donati}, J.-F., {Moutou}, C., {Bohlender}, D., {Catala}, C.,
  {Deleuil}, M., {Shkolnik}, E., {Collier Cameron}, A., {Jardine}, M.~M., \&
  {Walker}, G.~A.~H. 2009, \mnras, 398, 1383

\bibitem[{{Gaidos} {et~al.}(2000){Gaidos}, {G{\"u}del}, \& {Blake}}]{Gaidos00}
{Gaidos}, E.~J., {G{\"u}del}, M., \& {Blake}, G.~A. 2000, \grl, 27, 501

\bibitem[{{Gallet} \& {Bouvier}(2013)}]{Gallet13}
{Gallet}, F., \& {Bouvier}, J. 2013, \aap, 556, A36

\bibitem[{{Garraffo} {et~al.}(2013){Garraffo}, {Cohen}, {Drake}, \&
  {Downs}}]{Garraffo13}
{Garraffo}, C., {Cohen}, O., {Drake}, J.~J., \& {Downs}, C. 2013, \apj, 764, 32

\bibitem[{{Garraffo} {et~al.}(2015){Garraffo}, {Drake}, \&
  {Cohen}}]{Garraffo.etal:15}
{Garraffo}, C., {Drake}, J.~J., \& {Cohen}, O. 2015, \apjl, 807, L6

\bibitem[{{Gizon} \& {Rempel}(2008)}]{Gizon.Rempel:08}
{Gizon}, L., \& {Rempel}, M. 2008, \solphys, 251, 241

\bibitem[{{Hathaway}(2010)}]{Hathaway:10}
{Hathaway}, D.~H. 2010, Living Reviews in Solar Physics, 7, 1

\bibitem[{{Hollis} {et~al.}(1985){Hollis}, {Chin}, \& {Brown}}]{Hollis85}
{Hollis}, J.~M., {Chin}, G., \& {Brown}, R.~L. 1985, \apj, 294, 646

\bibitem[{{Holzwarth}(2005)}]{Holzwarth05b}
{Holzwarth}, G. 2005, ArXiv High Energy Physics - Phenomenology e-prints

\bibitem[{{Holzwarth} \& {Jardine}(2005)}]{Holzwarth05a}
{Holzwarth}, V., \& {Jardine}, M. 2005, \aap, 444, 661

\bibitem[{{Jardine} {et~al.}(2013){Jardine}, {Vidotto}, {van Ballegooijen},
  {Donati}, {Morin}, {Fares}, \& {Gombosi}}]{Jardine13}
{Jardine}, M., {Vidotto}, A.~A., {van Ballegooijen}, A., {Donati}, J.-F.,
  {Morin}, J., {Fares}, R., \& {Gombosi}, T.~I. 2013, \mnras, 431, 528

\bibitem[{{Jeffries}(2014)}]{Jeffries14}
{Jeffries}, R.~D. 2014, ArXiv e-prints

\bibitem[{{Johnstone} {et~al.}(2015){Johnstone}, {G{\"u}del}, {Brott}, \&
  {L{\"u}ftinger}}]{Johnstone15}
{Johnstone}, C.~P., {G{\"u}del}, M., {Brott}, I., \& {L{\"u}ftinger}, T. 2015,
  \aap, 577, A28

\bibitem[{{Kawaler}(1988)}]{Kawaler88}
{Kawaler}, S.~D. 1988, \apj, 333, 236

\bibitem[{{Kawaler}(1989)}]{Kawaler:89}
---. 1989, \apjl, 343, L65

\bibitem[{{Koenigl}(1991)}]{Koenigl91}
{Koenigl}, A. 1991, \apjl, 370, L39

\bibitem[{{Komm} {et~al.}(2011){Komm}, {Howe}, {Hill}, {Gonz{\'a}lez
  Hern{\'a}ndez}, \& {Haber}}]{Komm.etal:11}
{Komm}, R., {Howe}, R., {Hill}, F., {Gonz{\'a}lez Hern{\'a}ndez}, I., \&
  {Haber}, D. 2011, Journal of Physics Conference Series, 271, 012077

\bibitem[{{Komm}(1994)}]{Komm:94}
{Komm}, R.~W. 1994, \solphys, 149, 417

\bibitem[{{Lang} {et~al.}(2014){Lang}, {Jardine}, {Morin}, {Donati}, {Jeffers},
  {Vidotto}, \& {Fares}}]{Lang14}
{Lang}, P., {Jardine}, M., {Morin}, J., {Donati}, J.-F., {Jeffers}, S.,
  {Vidotto}, A.~A., \& {Fares}, R. 2014, \mnras, 439, 2122

\bibitem[{{Lanz} \& {Catala}(1992)}]{Lanz92}
{Lanz}, T., \& {Catala}, C. 1992, \aap, 257, 663

\bibitem[{{Lim} \& {White}(1996)}]{Lim96}
{Lim}, J., \& {White}, S.~M. 1996, \apjl, 462, L91

\bibitem[{{Marsden} {et~al.}(2011){Marsden}, {Jardine}, {Ram{\'{\i}}rez
  V{\'e}lez}, {Alecian}, {Brown}, {Carter}, {Donati}, {Dunstone}, {Hart},
  {Semel}, \& {Waite}}]{Marsden11a}
{Marsden}, S.~C., {Jardine}, M.~M., {Ram{\'{\i}}rez V{\'e}lez}, J.~C.,
  {Alecian}, E., {Brown}, C.~J., {Carter}, B.~D., {Donati}, J.-F., {Dunstone},
  N., {Hart}, R., {Semel}, M., \& {Waite}, I.~A. 2011, \mnras, 413, 1922

\bibitem[{{Matt} \& {Pudritz}(2004)}]{Matt.etal:04}
{Matt}, S., \& {Pudritz}, R.~E. 2004, \apjl, 607, L43

\bibitem[{{Matt} {et~al.}(2015){Matt}, {Brun}, {Baraffe}, {Bouvier}, \&
  {Chabrier}}]{Matt.etal:15}
{Matt}, S.~P., {Brun}, A.~S., {Baraffe}, I., {Bouvier}, J., \& {Chabrier}, G.
  2015, \apjl, 799, L23

\bibitem[{{Matt} {et~al.}(2012){Matt}, {MacGregor}, {Pinsonneault}, \&
  {Greene}}]{Matt12}
{Matt}, S.~P., {MacGregor}, K.~B., {Pinsonneault}, M.~H., \& {Greene}, T.~P.
  2012, \apj, 754, L26

\bibitem[{{Matt} {et~al.}(2010){Matt}, {Pinz{\'o}n}, {de la Reza}, \&
  {Greene}}]{Matt.etal:10}
{Matt}, S.~P., {Pinz{\'o}n}, G., {de la Reza}, R., \& {Greene}, T.~P. 2010,
  \apj, 714, 989

\bibitem[{{McComas} {et~al.}(2007){McComas}, {Velli}, {Lewis}, {Acton},
  {Balat-Pichelin}, {Bothmer}, {Dirling}, {Feldman}, {Gloeckler}, {Habbal},
  {Hassler}, {Mann}, {Matthaeus}, {McNutt}, {Mewaldt}, {Murphy}, {Ofman},
  {Sittler}, {Smith}, \& {Zurbuchen}}]{McComas07}
{McComas}, D.~J., {Velli}, M., {Lewis}, W.~S., {Acton}, L.~W.,
  {Balat-Pichelin}, M., {Bothmer}, V., {Dirling}, R.~B., {Feldman}, W.~C.,
  {Gloeckler}, G., {Habbal}, S.~R., {Hassler}, D.~M., {Mann}, I., {Matthaeus},
  W.~H., {McNutt}, R.~L., {Mewaldt}, R.~A., {Murphy}, N., {Ofman}, L.,
  {Sittler}, E.~C., {Smith}, C.~W., \& {Zurbuchen}, T.~H. 2007, Reviews of
  Geophysics, 45, 1004

\bibitem[{{Meibom} {et~al.}(2011){Meibom}, {Mathieu}, {Stassun}, {Liebesny}, \&
  {Saar}}]{Meibom11}
{Meibom}, S., {Mathieu}, R.~D., {Stassun}, K.~G., {Liebesny}, P., \& {Saar},
  S.~H. 2011, \apj, 733, 115

\bibitem[{{Mestel}(1968)}]{Mestel68}
{Mestel}, L. 1968, \mnras, 138, 359

\bibitem[{{Mestel}(1999)}]{Mestel99}
---. 1999, {Stellar magnetism, Oxford University Press, New York, USA}

\bibitem[{{Mestel} \& {Spruit}(1987)}]{Mestel87}
{Mestel}, L., \& {Spruit}, H.~C. 1987, \mnras, 226, 57

\bibitem[{{Michaud} \& {Charland}(1986)}]{Michaud86}
{Michaud}, G., \& {Charland}, Y. 1986, \apj, 311, 326

\bibitem[{{Mordvinov} \& {Yazev}(2013)}]{Mordvinov.Yazev:13}
{Mordvinov}, A.~V., \& {Yazev}, S.~A. 2013, Astronomy Reports, 57, 448

\bibitem[{{Morin} {et~al.}(2010){Morin}, {Donati}, {Petit}, {Delfosse},
  {Forveille}, \& {Jardine}}]{Morin10}
{Morin}, J., {Donati}, J.-F., {Petit}, P., {Delfosse}, X., {Forveille}, T., \&
  {Jardine}, M.~M. 2010, \mnras, 407, 2269

\bibitem[{{Oran} {et~al.}(2013){Oran}, {van der Holst}, {Landi}, {Jin},
  {Sokolov}, \& {Gombosi}}]{Oran13}
{Oran}, R., {van der Holst}, B., {Landi}, E., {Jin}, M., {Sokolov}, I.~V., \&
  {Gombosi}, T.~I. 2013, \apj, 778, 176

\bibitem[{{Owens} {et~al.}(2011){Owens}, {Crooker}, \&
  {Lockwood}}]{Owens.etal:11}
{Owens}, M.~J., {Crooker}, N.~U., \& {Lockwood}, M. 2011, Journal of
  Geophysical Research (Space Physics), 116, 4111

\bibitem[{{Parker}(1958)}]{Parker58}
{Parker}, E.~N. 1958, \apj, 128, 677

\bibitem[{{Phillips} {et~al.}(1995){Phillips}, {Bame}, {Barnes}, {Barraclough},
  {Feldman}, {Goldstein}, {Gosling}, {Hoogeveen}, {McComas}, {Neugebauer}, \&
  {Suess}}]{Phillips95}
{Phillips}, J.~L., {Bame}, S.~J., {Barnes}, A., {Barraclough}, B.~L.,
  {Feldman}, W.~C., {Goldstein}, B.~E., {Gosling}, J.~T., {Hoogeveen}, G.~W.,
  {McComas}, D.~J., {Neugebauer}, M., \& {Suess}, S.~T. 1995, \grl, 22, 3301

\bibitem[{{Poppenhaeger} {et~al.}(2012){Poppenhaeger}, {G{\"u}nther}, \&
  {Schmitt}}]{Poppenhaeger12}
{Poppenhaeger}, K., {G{\"u}nther}, H.~M., \& {Schmitt}, J.~H.~M.~M. 2012,
  Astronomische Nachrichten, 333, 26

\bibitem[{Powell {et~al.}(1999)Powell, Roe, Linde, Gombosi, \&
  Zeeuw}]{Powell99}
Powell, K.~G., Roe, P.~L., Linde, T.~J., Gombosi, T.~I., \& Zeeuw, D. L.~D.
  1999, Journal of Computational Physics, 154, 284

\bibitem[{{Rebull} {et~al.}(2004){Rebull}, {Wolff}, \&
  {Strom}}]{Rebull.etal:04}
{Rebull}, L.~M., {Wolff}, S.~C., \& {Strom}, S.~E. 2004, \aj, 127, 1029

\bibitem[{{R{\'e}ville} {et~al.}(2015){R{\'e}ville}, {Brun}, {Matt},
  {Strugarek}, \& {Pinto}}]{Reville.etal:15}
{R{\'e}ville}, V., {Brun}, A.~S., {Matt}, S.~P., {Strugarek}, A., \& {Pinto},
  R.~F. 2015, \apj, 798, 116

\bibitem[{{Schrijver} \& {Title}(2001)}]{Schrijver.Title:01}
{Schrijver}, C.~J., \& {Title}, A.~M. 2001, \apj, 551, 1099

\bibitem[{{Schuessler} \& {Solanki}(1992)}]{Schuessler.Solanki:92}
{Schuessler}, M., \& {Solanki}, S.~K. 1992, \aap, 264, L13

\bibitem[{{Semel}(1980)}]{Semel80}
{Semel}, M. 1980, \aap, 91, 369

\bibitem[{{Skumanich}(1972)}]{Skumanich72}
{Skumanich}, A. 1972, \apj, 171, 565

\bibitem[{{Soderblom} {et~al.}(1993){Soderblom}, {Stauffer}, {MacGregor}, \&
  {Jones}}]{Soderblom93}
{Soderblom}, D.~R., {Stauffer}, J.~R., {MacGregor}, K.~B., \& {Jones}, B.~F.
  1993, \apj, 409, 624

\bibitem[{{Sokolov} {et~al.}(2013){Sokolov}, {van der Holst}, {Oran}, {Downs},
  {Roussev}, {Jin}, {Manchester}, {Evans}, \& {Gombosi}}]{Sokolov13}
{Sokolov}, I.~V., {van der Holst}, B., {Oran}, R., {Downs}, C., {Roussev},
  I.~I., {Jin}, M., {Manchester}, IV, W.~B., {Evans}, R.~M., \& {Gombosi},
  T.~I. 2013, \apj, 764, 23

\bibitem[{{Solanki} {et~al.}(1997){Solanki}, {Motamen}, \&
  {Keppens}}]{Solanki97}
{Solanki}, S.~K., {Motamen}, S., \& {Keppens}, R. 1997, \aap, 325, 1039

\bibitem[{{Stauffer} \& {Hartmann}(1987)}]{Stauffer87}
{Stauffer}, J.~R., \& {Hartmann}, L.~W. 1987, \apj, 318, 337

\bibitem[{{T{\'o}th} {et~al.}(2012){T{\'o}th}, {van der Holst}, {Sokolov}, {De
  Zeeuw}, {Gombosi}, {Fang}, {Manchester}, {Meng}, {Najib}, {Powell}, {Stout},
  {Glocer}, {Ma}, \& {Opher}}]{Toth12}
{T{\'o}th}, G., {van der Holst}, B., {Sokolov}, I.~V., {De Zeeuw}, D.~L.,
  {Gombosi}, T.~I., {Fang}, F., {Manchester}, W.~B., {Meng}, X., {Najib}, D.,
  {Powell}, K.~G., {Stout}, Q.~F., {Glocer}, A., {Ma}, Y.-J., \& {Opher}, M.
  2012, Journal of Computational Physics, 231, 870

\bibitem[{{Ulrich} \& {Boyden}(2005)}]{Ulrich.Boyden:05}
{Ulrich}, R.~K., \& {Boyden}, J.~E. 2005, \apjl, 620, L123

\bibitem[{{van der Holst} {et~al.}(2014){van der Holst}, {Sokolov}, {Meng},
  {Jin}, {Manchester}, {T{\'o}th}, \& {Gombosi}}]{Vanderholst14}
{van der Holst}, B., {Sokolov}, I.~V., {Meng}, X., {Jin}, M., {Manchester}, IV,
  W.~B., {T{\'o}th}, G., \& {Gombosi}, T.~I. 2014, \apj, 782, 81

\bibitem[{{Vidotto} {et~al.}(2012){Vidotto}, {Fares}, {Jardine}, {Donati},
  {Opher}, {Moutou}, {Catala}, \& {Gombosi}}]{Vidotto12}
{Vidotto}, A.~A., {Fares}, R., {Jardine}, M., {Donati}, J.-F., {Opher}, M.,
  {Moutou}, C., {Catala}, C., \& {Gombosi}, T.~I. 2012, \mnras, 423, 3285

\bibitem[{{Vidotto} {et~al.}(2014{\natexlab{a}}){Vidotto}, {Gregory},
  {Jardine}, {Donati}, {Petit}, {Morin}, {Folsom}, {Bouvier}, {Cameron},
  {Hussain}, {Marsden}, {Waite}, {Fares}, {Jeffers}, \& {do
  Nascimento}}]{Vidotto14a}
{Vidotto}, A.~A., {Gregory}, S.~G., {Jardine}, M., {Donati}, J.~F., {Petit},
  P., {Morin}, J., {Folsom}, C.~P., {Bouvier}, J., {Cameron}, A.~C., {Hussain},
  G., {Marsden}, S., {Waite}, I.~A., {Fares}, R., {Jeffers}, S., \& {do
  Nascimento}, J.~D. 2014{\natexlab{a}}, \mnras, 441, 2361

\bibitem[{{Vidotto} {et~al.}(2014{\natexlab{b}}){Vidotto}, {Jardine}, {Morin},
  {Donati}, {Opher}, \& {Gombosi}}]{Vidotto14b}
{Vidotto}, A.~A., {Jardine}, M., {Morin}, J., {Donati}, J.~F., {Opher}, M., \&
  {Gombosi}, T.~I. 2014{\natexlab{b}}, \mnras, 438, 1162

\bibitem[{{Vidotto} {et~al.}(2011){Vidotto}, {Jardine}, {Opher}, {Donati}, \&
  {Gombosi}}]{Vidotto11}
{Vidotto}, A.~A., {Jardine}, M., {Opher}, M., {Donati}, J.~F., \& {Gombosi},
  T.~I. 2011, \mnras, 412, 351

\bibitem[{{Vidotto} {et~al.}(2010){Vidotto}, {Opher}, {Jatenco-Pereira}, \&
  {Gombosi}}]{Vidotto10}
{Vidotto}, A.~A., {Opher}, M., {Jatenco-Pereira}, V., \& {Gombosi}, T.~I. 2010,
  \apj, 720, 1262

\bibitem[{{Vilhu}(1984)}]{Vilhu:84}
{Vilhu}, O. 1984, \aap, 133, 117

\bibitem[{{Vilhu} \& {Walter}(1987)}]{Vilhu.Walter:87}
{Vilhu}, O., \& {Walter}, F.~M. 1987, \apj, 321, 958

\bibitem[{{Waite} {et~al.}(2011){Waite}, {Marsden}, {Carter}, {Hart}, {Donati},
  {Ram{\'{\i}}rez V{\'e}lez}, {Semel}, \& {Dunstone}}]{Waite11}
{Waite}, I.~A., {Marsden}, S.~C., {Carter}, B.~D., {Hart}, R., {Donati}, J.-F.,
  {Ram{\'{\i}}rez V{\'e}lez}, J.~C., {Semel}, M., \& {Dunstone}, N. 2011,
  \mnras, 413, 1949

\bibitem[{{Waite} {et~al.}(2015){Waite}, {Marsden}, {Carter}, {Petit},
  {Donati}, {Jeffers}, \& {Boro Saikia}}]{Waite15}
{Waite}, I.~A., {Marsden}, S.~C., {Carter}, B.~D., {Petit}, P., {Donati},
  J.-F., {Jeffers}, S.~V., \& {Boro Saikia}, S. 2015, \mnras, 449, 8

\bibitem[{{Wang} {et~al.}(2000){Wang}, {Lean}, \& {Sheeley}}]{Wang.etal:00}
{Wang}, Y.-M., {Lean}, J., \& {Sheeley}, Jr., N.~R. 2000, \grl, 27, 505

\bibitem[{{Weber} \& {Davis}(1967)}]{Weber67}
{Weber}, E.~J., \& {Davis}, Jr., L. 1967, \apj, 148, 217

\bibitem[{{Wood}(2006)}]{Wood06}
{Wood}, B. 2006, AGU Fall Meeting Abstracts, A5

\bibitem[{{Wood} {et~al.}(2014){Wood}, {M{\"u}ller}, {Redfield}, \&
  {Edelman}}]{Wood14}
{Wood}, B.~E., {M{\"u}ller}, H.-R., {Redfield}, S., \& {Edelman}, E. 2014,
  \apjl, 781, L33

\bibitem[{{Wood} {et~al.}(2002){Wood}, {M{\"u}ller}, {Zank}, \&
  {Linsky}}]{Wood02}
{Wood}, B.~E., {M{\"u}ller}, H.-R., {Zank}, G.~P., \& {Linsky}, J.~L. 2002,
  \apj, 574, 412

\bibitem[{{Wood} {et~al.}(2005){Wood}, {M{\"u}ller}, {Zank}, {Linsky}, \&
  {Redfield}}]{Wood05}
{Wood}, B.~E., {M{\"u}ller}, H.-R., {Zank}, G.~P., {Linsky}, J.~L., \&
  {Redfield}, S. 2005, \apjl, 628, L143

\bibitem[{{Wright} {et~al.}(2011){Wright}, {Drake}, {Mamajek}, \&
  {Henry}}]{Wright.etal:11}
{Wright}, N.~J., {Drake}, J.~J., {Mamajek}, E.~E., \& {Henry}, G.~W. 2011,
  \apj, 743, 48

\end{thebibliography}


\begin{table*}[ht]
\caption{Mass and Angular Momentum Loss Model Grid and Results}

\begin{tabular}{c c c|c  c c c}
\multicolumn{7}{c}{Mass Losses}\\
\hline
\multicolumn{3}{c|}{Case} & \multicolumn{4}{|c}{Mass Loss [$g/s$]}  \\ 
Dipole Strength & Spots Strength & Spots Polarity   & $30\deg$\tablenotemark{a} &
$60\deg$\tablenotemark{a} & $80\deg$\tablenotemark{a} & No Spots\\
\hline \hline 
  & {Observed}  &
original& 2.52e12&1.64e12 &1.51e12 & \multirow{4}{*}{3.68e12}\\
\cline{3-3} \cline{4-6}
10G Dipole&(Spots flux = 1.e24 Maxwell) & constant dipole comp & 2.6e12& 2.02e12& 1.7e12 & \\\cline{2-3}\cline{4-6}
(Flux = 2.27e23 Maxwell) &{Strong}& original &  4.84e12&& 1.56e12 & \\\cline{3-3} \cline{4-6}
&(Spots flux = 2.e24 Maxwell) & inverted & 5.16e12&& 1.91e12  & \\

\hline
 & {Observed}& original&
 3.34e12&2.7e12  &2.62e12 & \multirow{4}{*}{8.39e12} \\  \cline{3-3}\cline{4-6}
20G Dipole&(Spots flux = 1.e24 Maxwell) &  constant dipole comp & 3.28e12 &
3.23e12 & 2.95e12 & \\\cline{2-3} \cline{4-6}
(Flux = 4.54e23 Maxwell)&{Strong}& original &  4.62e12&& 2.62e12 & \\\cline{3-3} \cline{4-6}
&(Spots flux = 2.e24 Maxwell) & inverted &  4.91e12&&3.1e12 & \\

\multicolumn{7}{c}{} \\


\multicolumn{7}{c}{Angular Momentum Losses }\\
\hline
\multicolumn{3}{c|}{Case} & \multicolumn{4}{|c}{Angular Momentum Loss [$g \, cm^2/s^2$]}  \\ 
Dipole Strength & Spots Strength & Spots Polarity &$30\deg$\tablenotemark{a} &
$60\deg$\tablenotemark{a} & $80\deg$\tablenotemark{a} & No Spots\\

\hline \hline 
 & {Observed}  &
original&  7.11e30& 3.17e30& 3.33e30 &\multirow{4}{*}{8.37e30}\\ \cline{3-3} \cline{4-6}
10G Dipole&(Spots flux = 1.e24 Maxwell) & constant dipole comp  &8.57e30& 6.86e30& 5.76e30 &\\\cline{2-3}\cline{4-6}
(Flux = 2.27e23 Maxwell) &{Strong}& original  &  1.52e31&& 3.66e30 &\\\cline{3-3} \cline{4-6}
&(Spots flux = 2.e24 Maxwell) & inverted &  1.57e31&&6.57e30&\\

\hline
 & {Observed}& original&
 1.5e31& 1.05e31& 1.02e31&\multirow{4}{*}{2.24e31}\\  \cline{3-3}\cline{4-6}
20G Dipole&(Spots flux = 1.e24 Maxwell) &  constant dipole comp &1.65e31& 1.47e31&1.35e31& \\\cline{2-3} \cline{4-6}
(Flux = 4.54e23 Maxwell)&{Strong}& original   &1.98e31&& 1.16e31 & \\\cline{3-3} \cline{4-6}
&(Spots flux = 2.e24 Maxwell) & inverted &2.33e31&& 1.55e31& \\

\hline   
\end{tabular}

\tablenotetext{1}{The mean latitude of the spots in the magnetograms used to drive the simulations; $30\deg$ corresponds to the unshifted spots at their observed latitude.}

\label{table:aml}   
\end{table*}

\end{document}